\documentclass[%
 aip,
 jcp,
 amsmath,
 amssymb,
 preprint,%
floatfix
]{revtex4-1}

\usepackage[english]{babel}	
\usepackage[pdftex]{graphicx}
\usepackage{amsmath,amsfonts,amsthm,amssymb} 
\usepackage{changes}
\usepackage{hyperref}
\usepackage{float}
\usepackage{placeins}
\usepackage{textcomp}

\hypersetup{
pdftitle={Excitation dynamics in polyacene molecules on rare-gas clusters},
pdfauthor={Matthias Bohlen}
}

\begin{document}


\title{Excitation dynamics in polyacene molecules on rare-gas clusters} 



\author{Matthias Bohlen}
\email[]{matthias.bohlen@physik.uni-freiburg.de}
\author{Rupert Michiels}
\author{Moritz Michelbach}
\author{Selmane Ferchane}
\affiliation{Institute of Physics, University of Freiburg, Hermann-Herder-Str.\ 3, 79104 Freiburg, Germany}
\author{Michael Walter}
\affiliation{Institute of Physics, University of Freiburg, Hermann-Herder-Str.\ 3, 79104 Freiburg, Germany}
\affiliation{FIT Freiburg Center for Interactive Materials and Bioinspired Technologies, University of Freiburg, Georges-Köhler-Allee 105, 79110 Freiburg, Germany}
\affiliation{Fraunhofer IWM, MikroTribologie Centrum \textmu TC, Wöhlerstr.\ 11, 79108 Freiburg, Germany}
\author{Alexander Eisfeld}
\affiliation{Max Planck Institute for the Physics of Complex Systems, Nöthnitzer Str.\ 38, 01187 Dresden, Germany}
\author{Frank Stienkemeier}
\email[]{stienkemeier@uni-freiburg.de}
\affiliation{Institute of Physics, University of Freiburg, Hermann-Herder-Str.\ 3, 79104 Freiburg, Germany}

\date{December 21, 2021}

\noindent \textit{The following article has been accepted by The Journal of Chemical Physics. After it is published, it will be found at \url{https://doi.org/10.1063/5.0073503}.}\\

\begin{abstract}
Laser-induced fluorescence spectra and excitation lifetimes of anthracene, tetracene, and pentacene molecules attached to the surface of solid argon clusters have been measured with respect to cluster size, density of molecules and excitation density. Results are compared to previous studies on the same sample molecules attached to neon clusters. A contrasting lifetime behavior of anthracene on neon and argon clusters is discussed, and mechanisms are suggested to interpret the results. Although both neon and argon clusters are considered to be weakly interacting environments, we find that the excitation decay dynamics of the studied acenes depends significantly on the cluster material. Moreover, we find even qualitative differences regarding the dependence on the dopant density. Based on these observations, previous assignments of collective radiative and non-radiative decay mechanisms are discussed in the context of the new experimental findings.
\end{abstract}


\maketitle 


\section{Introduction}\label{sec:intro}
Over the last decades, polyacene dyes have transformed from materials of purely scientific research towards power horses of organic electronics. 
Their physical properties such as high and tunable charge carrier mobility,\cite{art:Nelson98,art:Li12,art:Lee19} above-unity quantum yields up to $200\,\%$ (through singlet fission)\cite{art:Xia16,art:Thompson13,art:Walker13} as well as their advantages for industrial utilization such as low cost\cite{art:Irimia-Vladu10,art:Zan12,art:Baude03} and room temperature processing,\cite{art:Nomura04,art:Lassnig15} or bio-compatibility\cite{art:Irimia-Vladu10} have lead to a large variety of applications, from high performance organic transistors,\cite{art:Lin97,art:Lin97a,art:Li12,art:Tan09,art:Khan11,art:Zan12} flexible displays,\cite{art:Sheraw02,art:Rogers01} and sensing applications\cite{art:Someya05,art:Khan11,art:Zan12} to efficient light harvesting in photovoltaic applications.\cite{art:Thompson13,art:Xia16}
It has already been shown that optical and electronic properties of these molecules are strongly dependent on their geometrical configuration towards each other and the environmental conditions.\cite{art:Peter80,art:Amirav82,art:Thorsmolle09,art:Lottner19} 
Moreover, for a multitude of these molecules, inter-molecular interactions and related effects such as superradiance,\cite{art:Burdett11,art:Mueller15,art:Camposeo10} singlet fission,\cite{art:Piland15,art:Burdett11,art:Korovina16,art:Walker13,art:Garoni17} and excimer formation\cite{art:Hofmann79,art:Kobayashi79,art:Chen10,art:Dover18,art:Sugino13,art:Ishii94} can alter the excitation dynamics and the resulting optical properties significantly.

Previous studies in our group have focussed on the excitation lifetime dependences of polyacenes, specifically anthracene (Ac), tetracene (Tc) and pentacene (Pc), attached to the surface of solid neon clusters, with respect to the number of molecules on the cluster and excitation density, aiming at a better understanding of collective radiative and non-radiative decay mechanisms and their impact on lifetime and quantum yield. 
The used cluster isolation spectroscopy method provides control of the number of molecules on the cluster surface.
A collective fluorescence rate enhancement for assemblies of these molecules was found with a strong dependence on the laser power.\cite{art:Mueller15, art:Izadnia17, diss:Izadnia2018} 
The findings were accredited to superradiance (SR), as well as a non-radiative deexcitation mechanism for dense ensembles.\cite{art:Mueller15} The nature of the non-radiative deexcitation channel, that was found to be similar for Tc and Pc, was later ascribed to be singlet fission (SF), where for Ac, a more complex pathway for the deexcitation process was proposed.\cite{art:Izadnia17, diss:Izadnia2018}

In these previous studies the focus was on the interactions between the molecules, and the influence of the neon clusters was not considered.
However, it is known that the presence of cluster can change the optical properties of molecules. 

The aim of the present study is to gain insight about the influence of the environment on the lifetimes.\cite{art:Mueller15,art:Izadnia17, diss:Izadnia2018}
Argon is an ideal material for this purpose, as it offers the next stronger interaction in the group of rare gases, but still provides a weakly interacting environment.
Also the temperature of $37\,$K is only slightly higher compared to neon ($10\,$K).\cite{art:Farges81}
As the changes of the properties of the cluster material are small, one expects that polyacenes on argon clusters behave very similarly as on neon. 
In the present study we use a similar approach as in our previous studies mentioned above.
The excitation lifetimes of Ac, Tc, and Pc attached to the surface of solid argon clusters are recorded, with respect to surface dopant density, cluster size and laser excitation density. 
Despite the precise control of experimental parameters, the complicated entanglements of experimental quantities necessitate utmost care regarding interpretations, particularly when comparing results from different studies. Therefore, we will direct our attention mainly on general trends, but not on the exact numerical quantities obtained.

The paper is organized as follows. In section~\ref{sec:met} we introduce the experimental and computational methods. In section \ref{sec:previous} we provide an overview of previous studies on the excitation decay dynamics of polyacenes attached to neon clusters. In section~\ref{sec:res} we present the observations for the argon system. In section~\ref{sec:dis} we compare the argon results to results on neon clusters to obtain insight into the changes induced by the interaction of dopant molecule and environment. Finally, in section~\ref{sec:concl}, we summarize our findings and give a brief outlook.

\section{Methods}\label{sec:met}
\begin{figure*}
\centering
\includegraphics[width=17cm]{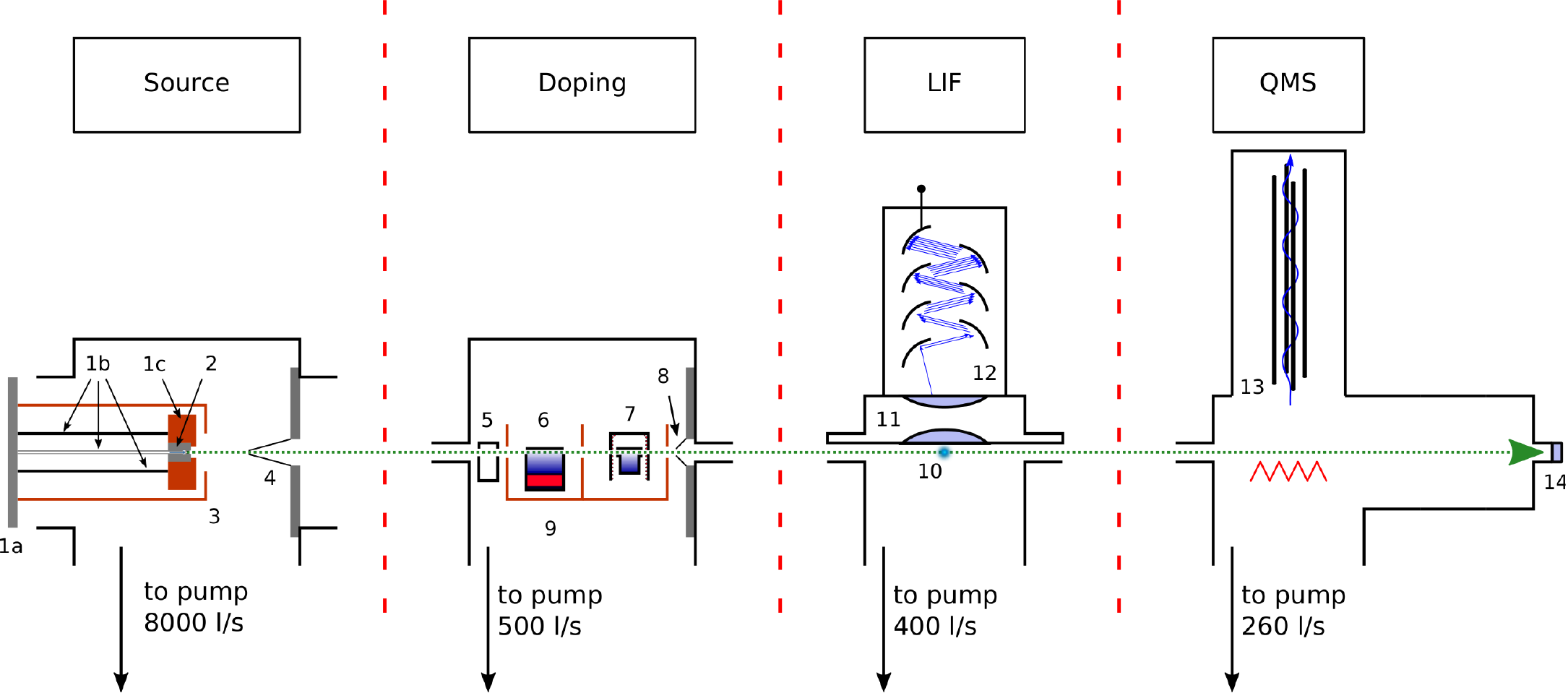}
\caption{Schematic of the experimental apparatus. 1: Source module (a: main flange, b: separator rods, gas supply, electrical connections, LN$_2$ supply, c: copper valve mount), 2: pulsed valve, 3: water cooled copper shield, 4: skimmer, 5: gas doping cell, 6: cartridge heated vaporizer cell, 7: radiative heating vaporizer cell, 8: skimmer, 9: water cooled copper shield, 10: laser interaction point, 11: lens assembly, 12: photomultiplier tube (PMT), 13: quadrupole mass spectrometer, 14: rear window. The dimensions are not to scale.}
\label{fig:setup}
\end{figure*}
\paragraph{Experimental setup:}\label{sec:met_setup}
A schematic sketch of the experimental setup is shown in Fig.~\ref{fig:setup}. The clusters are produced in the source chamber by a supersonic expansion from a pulsed valve (Even-Lavie valve\cite{art:Even00,art:Luria11}).
The pressurized argon gas (stagnation pressure $50\,$bars, purity $\geq 99.999\,\%$) is expanded through a front orifice of $60$ or $100\,$\textmu m in diameter with driver pulse lengths of $22-24\,$\textmu s at $200\,$Hz repetition rate for the laser-induced fluorescence (LIF) spectra and $50$ to $100\,$Hz for all other measurements.
The valve is mounted onto a home-built liquid nitrogen (LN$_2$) cooling stage with adaptive counter heating (controller model Omega CN 7600). This source assembly allows for temperatures down to $100\,$K to be reached. The chamber is pumped by an oil diffusion pump ($8000\,$l/s) backed by a roots blower and a rotary pump, maintaining a vacuum pressure below $10^{-4}\,$mbar during valve operation. A water-cooled shield protects the valve from the pump heat and reduces condensation of pump oil on the source module. 
The cluster beam proceeds to the next chamber through a conical shaped skimmer with $2\,$mm opening diameter.
The doping chamber houses a gas doping cell for gaseous sample materials and 2 slots for different types of vaporizer cells for solid or liquid samples. In this study, a cartridge heated vaporizer cell of $25\,$mm in length was used to evaporate the powder samples for the gas-phase molecules to collide with and stick to the clusters (pick-up method\cite{art:Gough85}). The doping chamber is pumped by a turbo molecular pump ($500\,$l/s), providing pressures in the $10^{-6}$ to $10^{-7}\,$mbar range. 
After passing through a second skimmer orifice at the end of the doping chamber, the beam of clusters with attached dopant molecules enters the LIF chamber, where the dopants can be excited by pulsed laser light. 
For the excitation of the dopant molecules, we use two pulsed dye laser systems (Sirah Cobra, one with optional frequency doubling unit), that are pumped by a frequency doubled and a frequency tripled Nd:YLF solid state laser (Edgewave Innoslab IS8II-E, Innoslab IS8III-E), respectively, covering a wavelength range of about $12500 - 36000\,$cm$^{-1}$. For this study, the frequency doubled system was used with Pyridine 2 as laser dye and the KDP (potassium dihydrogen phosphate) crystal frequency doubling unit ($\approx 27000 - 28300\,$cm$^{-1}$) for Ac, and the frequency tripled system was used with Coumarin 47 laser dye ($\approx 21450 - 22500\,$cm$^{-1}$) for Tc and with Coumarin 153 laser dye ($\approx 17850 - 19300\,$cm$^{-1}$) for Pc. 
Upon laser excitation, fluorescence or scattered light is detected by a photomultiplier tube (PMT, model Hamamatsu R 5600-U-01) through a lens assembly to maximize the detection solid angle. The laser beam enters the chamber at right angle relative to both cluster beam axis and PMT direction. A baffle system guarantees straylight suppression. The LIF chamber is pumped by a turbo molecular pump ($400\,$l/s), keeping the vacuum at low to medium $10^{-7}\,$mbars.
Lastly, the cluster beam proceeds to the QMS chamber, housing an EXTREL MAX1000 quadrupole mass spectrometer, allowing for cluster beam characterization. That chamber is pumped by another turbomolecular pump ($260\,$l/s), providing operational pressures in the low $10^{-8}\,$mbar range.

The clusters are created by a supersonic expansion from a pulsed valve, which will lead to a Poissonian distribution of cluster sizes. 
We estimate the mean cluster size, using the Hagena scaling laws, extended for large argon clusters by Dorchies et al.\cite{art:Hagena92,art:Buck96,art:Dorchies03} 
For the cluster sizes used in this work (radius $\approx 4$ nm to $\approx 10$ nm) the clusters consist of $\approx 10^4$ to $\approx 10^5$ atoms, which is well beyond the cluster sizes, where structural transitions are reported (hundreds to few thousands of atoms).\cite{art:Lee87} Therefore, the clusters are considered to be in a bulk-like fcc structure.
The cluster temperature and therefore the resulting temperature of the dopants on the cluster surface can be estimated\cite{book:Pauly2000} with the heat of sublimation of $\Delta u\approx 8000\,$J/mole\cite{art:Ferreira08} as $T=0.04\frac{\Delta u}{k_B}\approx 38.5\,$K, which is in good agreement with the experimental value by Farges et al.\cite{art:Farges81} of $37\,\pm\,5\,$K. 

The dopant materials anthracene (Sigma-Aldrich, purity $99\,\%$), tetracene (Sigma-Aldrich, purity $99.9\,\%$) and pentacene (Sigma-Aldrich, purity $99.9\,\%$) were used without further purification.
The dopant partial pressures $P$ in the vapor cells are calculated by the August equation $\log(P) = A-B/T$, using the temperatures of the vapor cell and literature values for $A$ and $B$ (anthracene: Ref.~\citenum{techrep:Suuberg97}, tetracene and pentacene: Ref.~\citenum{book:Stephenson1987}). The vapor cell dopant density $\rho$ is calculated by $\rho = P/k_BT$. The clusters are assumed to be approximately spherical. Then the mean number of dopants on a cluster $N$ can be estimated with the length of the vapor cell $l$ and the cluster radius $R$ to be $N = l\rho R^2 \pi$, and with this, the dopant density on the surface is $\sigma = l\rho/4$. The probability for picking up $k$ dopant molecules on a cluster is determined by the Poissonian distribution $P_k(\rho)$. The mean number of molecules per cluster $N(\rho)$ is then given by
\begin{equation}\label{eqn:mean_dopant_number}
N = \sum_{k=0}^\infty P_k(\rho)\cdot k 
\end{equation}

We further assume the dopant molecules to follow a statistical distribution regarding their positions. A mean inter-molecular distance $d$ can then be calculated by $d = 2R \sin\left(\langle\Theta\rangle/2\right)$ with the mean angular distance of neighboring molecules $\langle\Theta\rangle$, as given in Ref.~\citenum{art:Scott89}.

\paragraph{Extraction of lifetimes from time resolved fluorescence:}
For the analysis of the time evolution of the signal, the LIF signal was deconvoluted from a time-resolved Rayleigh scattering signal of a bare cluster beam (without doping), which was obtained with the laser tuned to the measurement wavenumber for Pc ($\nu=18 000\,$cm$^{-1}$) and a mean cluster radius of $9.3\,$nm. 
The Rayleigh signal consists of a Gaussian-shaped laser peak of about $8\,$ns total width with a weak echo, about $7.5\,$ns later. The laser that was used for obtaining the Rayleigh signal (Innoslab IS8-IIIE) was used for all Tc and Pc measurements. 
For the Ac measurements, the other laser (Innoslab IS8-IIE) was used with subsequent frequency doubling. Due to low laser power density for this laser, we did not obtain any Rayleigh signal at the Ac measurement conditions.
Therefore, the Rayleigh signal as obtained for Pc measurement conditions was used. The pulse width of the laser used for Ac was slightly longer than the one for Tc and Pc, with FWHM of the pulse widths of $5.3$ (for both fundamental and second harmonic) and $3.9\,$ns, respectively. 
This difference is in the range of the fluctuations of the extracted lifetimes and does not change the interpretation. 
A monoexponential fit model was applied in good agreement with the deconvoluted signal throughout all measurements.
A biexponential model was also tested, but did not provide any improvement on the fitting. 
The results of the lifetime fits are subject to fluctuations, which are mostly in the range of about $\pm\,1\,$ns for the measurements with strong signal levels (i.e., at maximum laser excitation and high dopant densities), and about $\pm\,3\,$ns for the weakest signal conditions (i.e., low dopant density and low laser excitation density).

\paragraph{Quantum yield (QY) measurements:}
Since we can not measure the total absorption, we can not obtain {\it absolute} quantum yields. Using the total fluorescence (recorded by the PMT within the time interval of $200\,$ns after the laser pulse), and the vapor cell dopant density, which is for a constant cluster size proportional to the number of dopants on the cluster surface, the relative LIF per dopant is calculated as total fluorescence divided by dopant density. 
Under the assumption of constant absorption, the LIF per dopant is equal to a relative quantum yield. 
As we did not observe any sign of change in the absorption of Tc and Pc for varying densities of dopants on the cluster surface (cf.~Fig.~\ref{fig:spectra}), we consider the LIF per dopant to be equal to the {\it relative quantum yield}, which we denote by QY. 
For Ac, the LIF spectra, shown in Fig.~\ref{fig:spectra}a, exhibit a change upon variation of the surface dopant density, hence, the assumption of constant absorption does not hold for this case. Furthermore, due to low signal intensities and a resulting low signal-to-noise ratio, the LIF per dopant will not be considered.

\paragraph{Calculation of transition dipoles:}
We also make use of theoretical dipole elements and oscillator strengths calculated for Ac, Tc and Pc in the gas phase within time-dependent density functional theory (TDDFT) as implemented in the GPAW package\cite{art:Mortensen05,art:Enkovaara10,art:Walter08}. Numerical settings are chosen as described in Stauffert et al.\cite{art:Stauffert19}

\section{Summary of our previous studies on neon clusters}\label{sec:previous}
Since the present work focuses on the comparison between argon and neon clusters, a brief overview of the main results from our previous studies of acenes on neon clusters \cite{art:Mueller15, art:Izadnia17, diss:Izadnia2018} shall be given.
The study by Izadnia et al.\cite{art:Izadnia17} presented excitation lifetimes and LIF per dopant for Ac, Tc and Pc. 
As explained in Sec.~\ref{sec:met} we treat the LIF per dopant to be equal to the relative QY.
It was found that Tc and Pc exhibit a nearly identical decrease in lifetime upon increasing number of dopants.
The lifetime decreases continuously from approximately $32\,$ns at a surface dopant density of $0.003/\mathrm{nm}^2$ to approximately $2\,$ns at a density of $0.15/\mathrm{nm}^2$. Remarkably, the QY is constant for low surface dopant densities up to about $0.02/\mathrm{nm}^2$ (the border between region I and II); for larger densities it decreases and becomes virtually zero for surface dopant densities around $0.1/\mathrm{nm}^2$. The lifetime reduction was only weakly laser power dependent for several investigated inter-dopant distances (cf.\ Fig.~1c in Ref.~\citenum{art:Izadnia17}). 
It was suggested that superradiance is responsible for the lifetime reduction at constant QY. For the large number of dopants where the QY decreases, it was concluded that non-radiative effects become important in this region.
By considering the dependence of the lifetime on the dopant density and the laser power, it was concluded that SF contributes strongly to the loss of lifetime and QY for large dopant density as it leads to a loss of QY for molecules in close vicinity, and is very fast compared to the normal or the SR enhanced fluorescence. 
For Ac, where SF is energetically forbidden ($S_0\leftarrow S_1: 3.43\,$eV\cite{art:Lambert84}, 2$\cdot$($S_0\leftarrow T_1$): $3.66\,$eV\cite{art:Zalesskaya02}, for unbound molecules), a slightly different, but still similar to Tc an Pc, reduction of lifetime as function of the number of dopants was found.
However, here the QY starts decreasing already for small numbers of dopants. 
Furthermore, for Ac the lifetime reduction was strongly laser power dependent.
This rather complex dynamics was ascribed to excimer delayed fluorescence.\cite{diss:Izadnia2018} 

The study by Müller et al.\cite{art:Mueller15} further elucidated the cluster size dependence of the excitation dynamics for Tc and Pc. 
There, the lifetime reduction at the same number of dopants was found to depend strongly on the cluster size;
the smaller the cluster, the larger the lifetime reduction.
From this observation it was concluded that the lifetime reduction is stronger for smaller mean distances between the molecules.
Consequently, the lifetime has been studied as a function of the mean molecular distance and it was found that for the same mean distance there is only a weak dependence on cluster size.
For mean inter-dopant distances above $1.5\,$nm, larger clusters exhibit slightly larger lifetimes. For the smallest distances, all lifetimes become virtually identical (cf. Fig.~4 of Ref.~\citenum{art:Mueller15}).

In the previous investigations LIF spectra were recorded for all three acenes (Ac and Pc in Ref.~\citenum{diss:Izadnia2018}, Tc in Ref.~\citenum{art:Mueller15}) covering the $0_0^0$ transition and the first vibration. It was found that these spectra are independent of the surface dopant density.

\section{Experimental observations}\label{sec:res}
In this section we present results that are obtained for different cluster sizes, different vapor cell dopant densities and different laser powers.
Each combination of cluster size and vapor cell density gives a specific number of dopants on the cluster and a corresponding surface dopant density (when there are more than one molecule on the cluster).
The laser power influences the number of excited molecules which plays for example a role for superradiance.
The surface dopant density determines the distances between the molecules and therefore the mutual interactions, which are for example relevant for singlet fission, and also for superradiance.

\begin{figure}
    \centering
    \includegraphics[width=8.2cm]{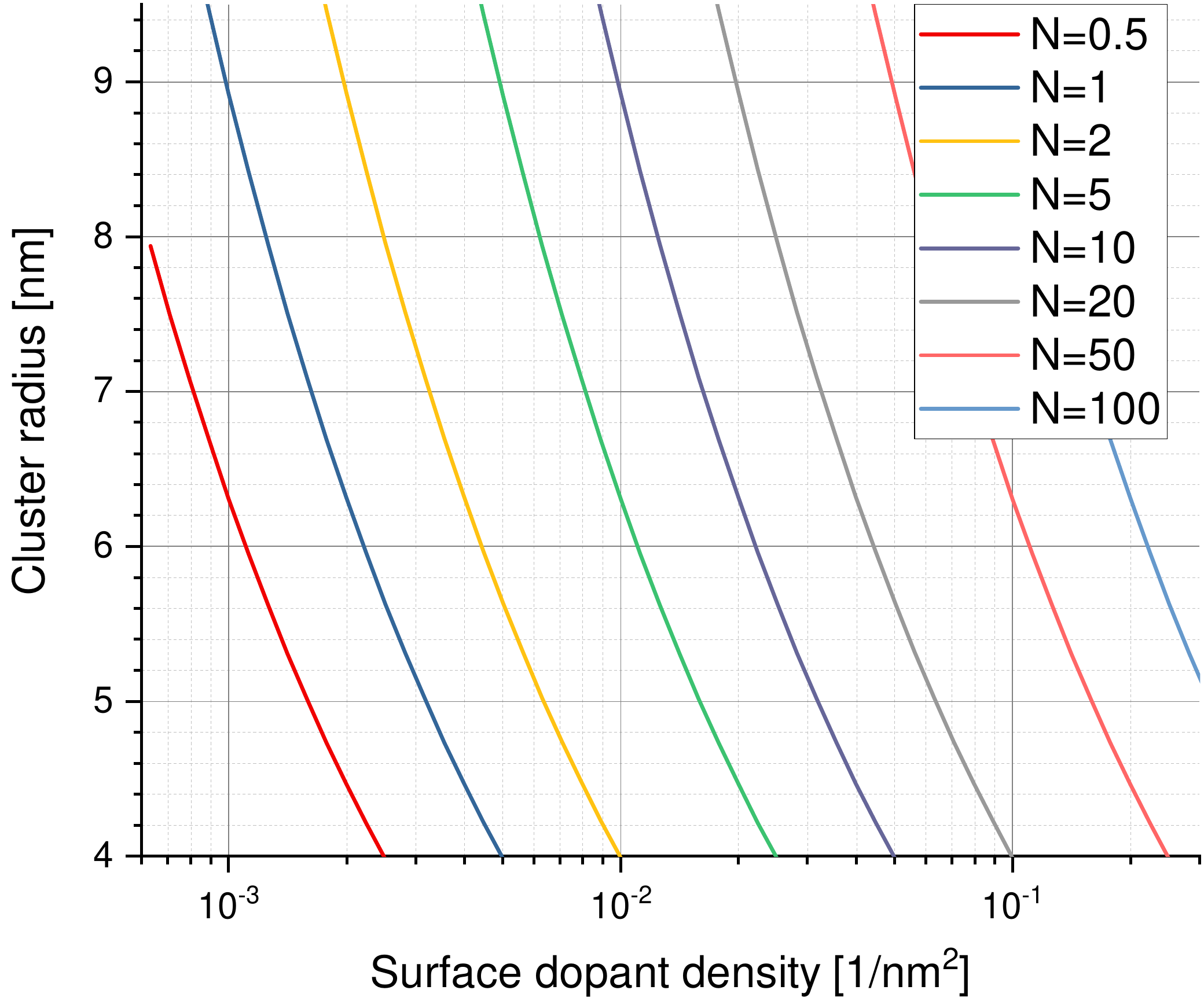}
    \caption{Estimated average numbers of dopants per cluster $N$ as a function of surface dopant density and cluster radius, as introduced in sec.~\ref{sec:met}a, Eq.~(\ref{eqn:mean_dopant_number}).}
    \label{fig:number_vs_density_radius}
\end{figure}

We convert the vapor cell dopant density to surface dopant density, as described in Sec.~\ref{sec:met}, and present our results in terms of this quantity and the mean cluster radius.
These two quantities then provide the number of dopants on the clusters.
We will often present plots where we show a certain quantity as function of the surface dopant density for fixed cluster sizes.
Here, one has to keep in mind that along such a function the mean number of molecules on the cluster changes proportionally to the surface dopant density and that for small clusters and low dopant densities one can have a single molecule on the cluster (and then surface density or inter-dopant distance are no longer meaningful numbers).
To facilitate the interpretation of the following results, in Fig.~\ref{fig:number_vs_density_radius} we present the mean number of molecules as function of surface dopant density and cluster radius. 
The range of values correspond to that used in the present study. 
Following the contour lines for the given absolute numbers of dopants, one can see how the number of dopants corresponds to the surface dopant density when the cluster size is varied. While for the smallest cluster size, one needs a surface dopant density of $10^{-2}/\mathrm{nm}^2$ to obtain two molecules on average on the clusters,
for the largest cluster sizes, this average number of dopants is reached already at lower surface dopant densities around $2.5\cdot 10^{-3}/\mathrm{nm}^{2}$. At the lowest surface dopant density of $10^{-3}/\mathrm{nm}^2$ there is on average only one molecule or less on the cluster, independent of the cluster size.
This means that the probability to have 2 molecules attached to the cluster is much smaller than the probability to have only one dopant molecule (or none).

\FloatBarrier

\subsection{LIF spectra}
\begin{figure}[tbp]
\centering
\includegraphics[width=8.2cm]{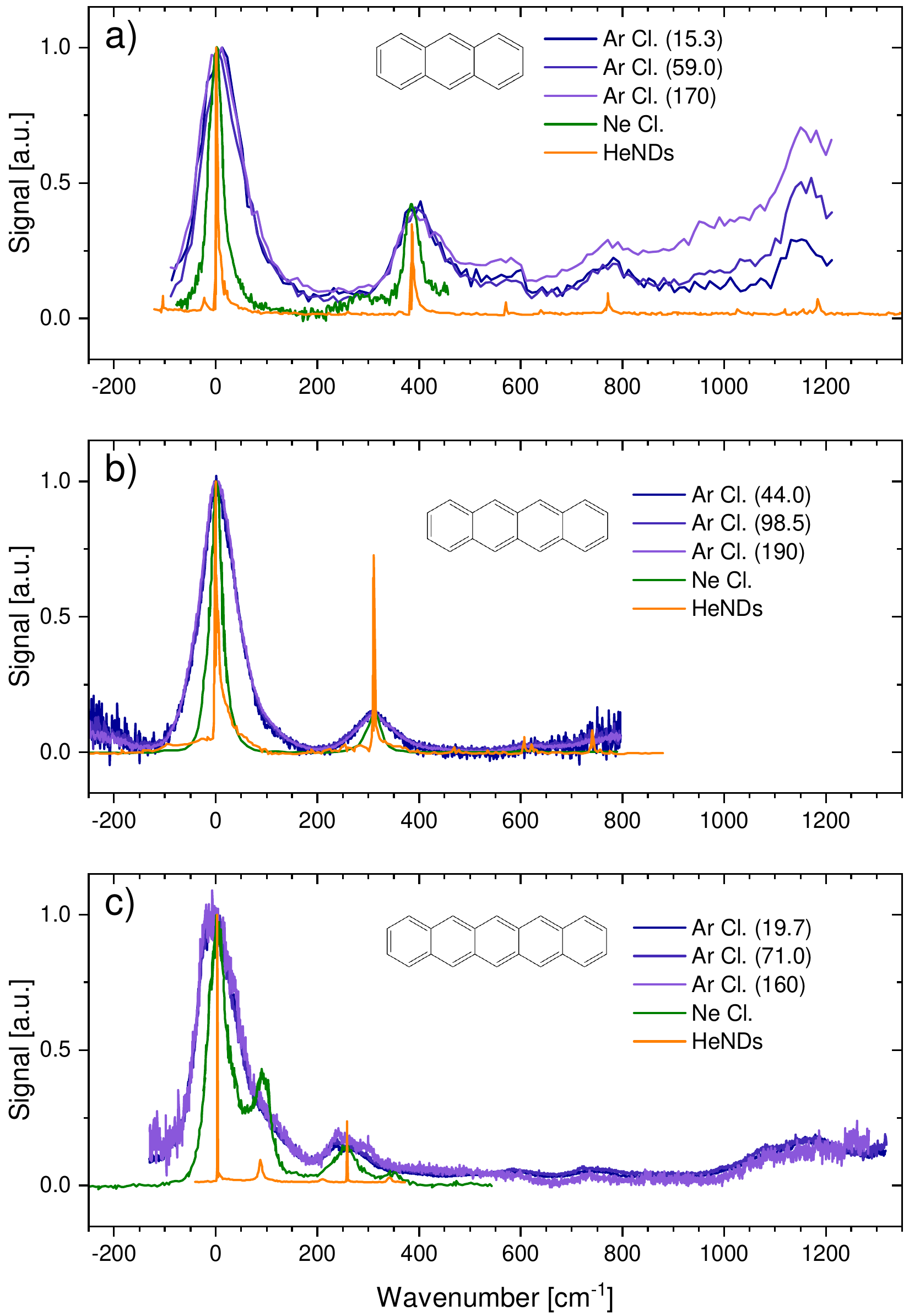}
\caption{LIF spectra of a) Ac, b) Tc, and c) Pc on the surface of argon clusters (at three different dopant densities), neon clusters, and embedded inside helium nanodroplets.
All spectra are normalized to the maximum of the respective $0_0^0$ transition. 
For all spectra the zero of the wavenumber axis is located at the maximum of the 0-0 transition.
For details on transition energies and shifts, see Tab.~\ref{tab:00positions}. The values in the legend give the corresponding surface dopant density in units of $10^{-3}\,$m$^{-2}$ at which the spectrum was recorded.
The spectra of acenes on neon clusters or embedded in HeNDs are taken from the following resources: Ac in He from Ref.~\citenum{art:Pentlehner10}, Ac on Ne from Ref.~\citenum{diss:Izadnia2018} , Tc in He from Ref.~\citenum{art:Hartmann01}, Tc on Ne from Ref.~\citenum{art:Mueller15}, Pc in He from Ref.~\citenum{art:Hartmann01}, Pc on Ne from Ref.~\citenum{diss:Izadnia2018}. The number of dopants on the neon clusters did not induce significant changes in the spectra in the investigated ranges of dopant numbers, ranging from monomers to tens of dopant molecules. The laser excitation density was $44\,$kW/cm$^2$.
}
\label{fig:spectra}
\end{figure}
Differences and similarities between neon and argon clusters can already be seen from LIF spectra.
In Fig.~\ref{fig:spectra}, we show such LIF spectra of Ac, Tc and Pc on argon clusters for three dopant densities along with spectra on the surface of solid neon clusters and spectra where the acenes are embedded inside helium nanodroplets (HeNDs), for comparison.

Laser excitation is from the electronic ground state $S_0$ ($^1 A_g$) to the first electronic excited state $S_1$ ($^1 B_{2u}$ for Ac, Tc, and Pc)\cite{art:Yang16} (Q band).

All spectra exhibit the purely electronic transition, a prominent first vibration (with wavenumber around $200-400\,$cm$^{-1}$), and several weakly active higher vibrational modes. 
For better comparison we have shifted all spectra such that the maximum of the spectra are at the same wavenumber, which we choose as the zero, i.e., we express wavenumbers relative to the purely electronic transition, which we denote by $0^0_0$. 
The actual wavenumbers of the $0^0_0$ transition are given in Tab.~\ref{tab:00positions}.
The signal intensity is normalized to the peak of the $0^0_0$ transition.

We make the following observation: for all three acenes there is an increasing redshift of the purely electronic transition and an increasing broadening of the peaks when going from HeND to neon and then to argon (see Tab.~\ref{tab:00positions}).
However, the relative positions and intensities of the vibrational transitions remain mostly unaffected.
For Pc on neon, a butterfly mode constitutes prominently\cite{diss:Izadnia2018,art:Stauffert19} at about $85\,$cm$^{-1}$, in agreement with results from a seeded beam experiment\cite{art:Amirav80} of $77\,$cm$^{-1}$. 
For argon this transition is visible only as a weak shoulder.
The much higher intensity of the first vibration of Tc in HeNDs relative to the 0-0 transition, compared to neon and argon clusters, is most likely a saturation effect due to high laser power used in Ref.~\citenum{art:Hartmann01}, from where the spectra are taken.
For Tc on argon a significant increase on the red side of the $0_0^0$ transition is observed around and below $-200\,\mathrm{cm}^{-1}$.
We assign this to complex formation of Tc with residual water (a similar observation and assignment was made for Tc embedded in HeNDs\cite{art:Lindinger06}).

The radii of the argon clusters used for recording the spectra are $5.6\,$nm (Ac/Tc) and $7.2\,$nm (Pc). With these cluster sizes, even for the lowest dopant densities the clusters are doped with more than one dopant on average (Ac: 1.5, Tc: 4.3, Pc: 3.2). 
The mean next-neighbor inter-dopant distances were around $8\,$nm for Ac and Pc and $5\,$nm for Tc for the lowest dopant densities.
For the highest dopant densities used, the mean number of dopants per cluster was 17, 19, and 26 for Ac, Tc, and Pc, respectively, and the mean distance between neighboring molecules was around $2.5\,$nm. 

One sees that for the range of measured wavenumbers the LIF spectra of Pc and Tc on argon are independent from the doping density.
A similar observation was made for the spectra on neon clusters. 
However, for Ac on argon we see that with increasing dopant density (and thus increasing number of molecules on the clusters) the relative weight of the high-lying vibrations is increased.
These observations show that the number of molecules and the surface dopant density can have a strong effect even for the LIF spectra.
We discuss this further in section \ref{sec:Dis_LIF}.

Furthermore, for Tc we have recorded the LIF spectra for different cluster sizes (radii from 4.1\,nm to 7.2\,nm, surface dopant density $\approx 0.1/\,\mathrm{nm}^{2}$, laser excitation density $44\,$kW/cm$^2$). The variation of the cluster size in the given range induced neither line shifts nor changes of peak profiles (not shown).

\begin{table*}
  \centering
    \begin{tabular}{l||c|c|c|c|c|c|c|c|c|c|c}
      \phantom{KKK}& \multicolumn{2}{c|}{On Ar cluster} & \multicolumn{2}{c|}{On flat Ar surface} & \multicolumn{2}{c|}{On Ne cluster} & \multicolumn{2}{c|}{On flat Ne surface} & \multicolumn{2}{c|}{In HeNDs} & \multicolumn{1}{c}{In vacuum} \\
      & \multicolumn{1}{c|}{$\nu$} & \multicolumn{1}{c|}{$\Delta \nu$} & \multicolumn{1}{c|}{$\nu$} & \multicolumn{1}{c|}{$\Delta \nu$} & \multicolumn{1}{c|}{$\nu$} & \multicolumn{1}{c|}{$\Delta \nu$} & \multicolumn{1}{c|}{$\nu$} & \multicolumn{1}{c|}{$\Delta \nu$} & \multicolumn{1}{c|}{$\nu$} & \multicolumn{1}{c|}{$\Delta \nu$} & \multicolumn{1}{c}{$\nu$} \\ \hline
    Ac    & 27084 & 611   &       &       & 27540\cite{diss:Izadnia2018} & 155   &       &       & 27622\cite{art:Pentlehner10} & 73    & 27695\cite{art:Lambert84} \\
    Tc    & 21705 & 692   &       &       & 22208\cite{art:Mueller15} & 189   &       &       & 22293\cite{art:Hartmann01} & 104   & 22397\cite{art:Dick91} \\
    Pc    & 17981 & 668   & 17864\cite{art:Halasinski00} & 785   & 18447\cite{diss:Izadnia2018} & 202   & 18426\cite{art:Halasinski00} & 223   & 18545\cite{art:Hartmann01} & 104   & 18649\cite{art:Heinecke98} \\
    \end{tabular}
  \caption{Experimental values of maxima of the $S_0^0\leftarrow S_1^0$ transitions of polyacenes in various media and respective redshifts $\Delta \nu$ with respect to the $0_0^0$ transition of the free molecule. Data for argon clusters are from this study, all others from respective references. All values in cm$^{-1}$.}
  \label{tab:00positions}%
\end{table*}

\subsection{Excitation lifetimes}
\begin{figure}[tbp]
\centering
\includegraphics[width=8.cm]{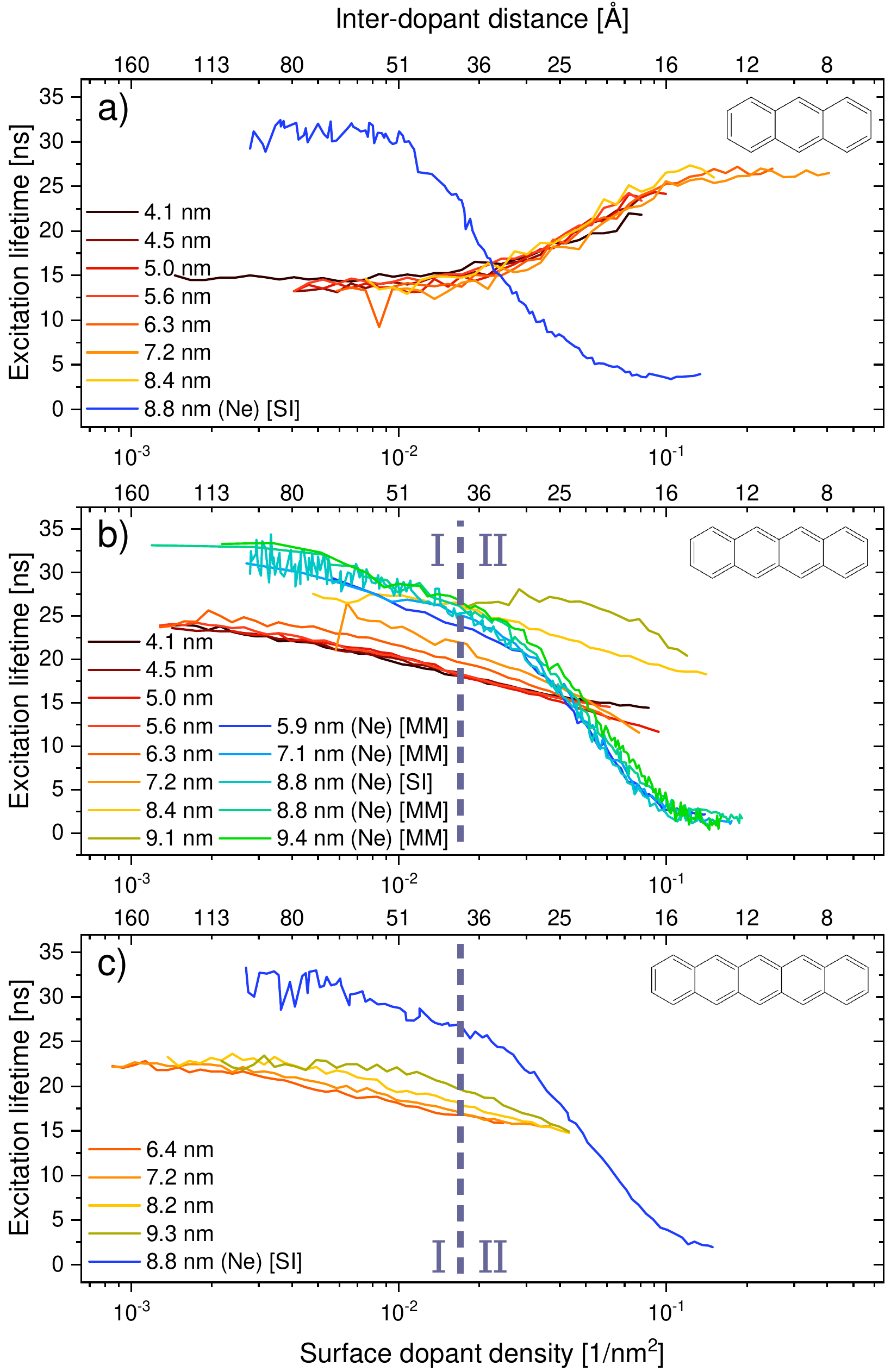}
\caption{Excitation lifetimes of a) Ac, b) Tc and c) Pc on Argon and Neon as a function of surface dopant density. Neon data labeled by [MM] and [SI] are taken from Refs.~\citenum{art:Mueller15} and \citenum{art:Izadnia17}, respectively. Note, that the laser power density differs: (Neon: Ac: $3\,$kW/cm$^2$, Tc and Pc by [SI]: $8\,$kW/cm$^2$, Tc by [MM]: $13.9\,$kW/cm$^2$.
Argon: Ac: $1.0\,$kW/cm$^2$, Tc and Pc: $44\,$kW/cm$^2$.)
The lower abscissae give the surface dopant density, the upper abscissae give the respective inter-dopant distance.
}
\label{fig:lifetimes_clustersize}
\end{figure}

In Fig.~\ref{fig:lifetimes_clustersize}, the excitation lifetimes of Ac, Tc and Pc are shown as functions of the dopant density.
The red to yellow colors show the lifetimes of acenes on argon clusters (different color indicating different cluster radii, as provided in the legend). 
The blue to green colors show data of the respective acene molecules on the surface of neon clusters, taken from Refs.~\citenum{art:Mueller15} and \citenum{art:Izadnia17} (with number of dopants converted to dopant density).
In our measurements on argon, the excitation laser was tuned to $27084\,$cm$^{-1}$ (Ac), $21705\,$cm$^{-1}$ (Tc) and $18000\,$cm$^{-1}$ (Pc), respectively. 
These wavenumbers correspond to the respective maxima of the LIF shown in Fig \ref{fig:spectra}.
The power densities shown here are $1.0\,$kW/cm$^{2}$ (pulse energy $0.33\,$\textmu J) in the case of Ac, and $44\,$kW/cm$^{2}$ ($11.1\,$\textmu J) for Tc and Pc. 
We comment on the dependence on laser power below.
Per data point, about 8000 (Ac, Tc) or 4000 (Pc) laser shots were averaged. \

The lifetime measurements on argon show an unexpected behavior, most remarkable for Ac, which behaves qualitatively different compared to all other cases.
For Ac on argon the lifetime {\it increases} upon {\it increasing} the dopant density. 
In all other cases, particularly including Ac on neon, the lifetime {\it decreases} upon increasing the dopant density. Hence, we have evidence that small changes in the cluster material can substantially change the decay dynamics.
We will now discuss all lifetime dependencies shown in Fig.~\ref{fig:lifetimes_clustersize} in more detail.

\paragraph{Anthracene:}
As already mentioned above, Ac on argon shows a qualitatively different behavior than on neon.
We would like to emphasize that there is a strong lifetime reduction at low dopant densities when going from neon to argon ($\approx30$\,ns on neon; $\approx15$\,ns on argon).
At these low dopant densities there are on average around 0.3 molecules on the smallest cluster and around 3 on the largest clusters. 
Around the dopant densities where on neon a decrease of lifetime happens, on argon an increase occurs.
For large dopant densities the lifetime of Ac on argon becomes comparable to the lifetime on neon at low dopant densities.
The excitation lifetimes recorded for Ac on argon were independent from the cluster size.

\paragraph{Tetracene:}
Tc shows a lifetime reduction towards higher doping in both the neon and the argon environment (Fig.~\ref{fig:lifetimes_clustersize}b).
Recapitulating the results from the neon system,\cite{art:Mueller15, art:Izadnia17} the lifetime function can be divided into two regions, marked as I/II in Fig.~\ref{fig:lifetimes_clustersize}b.
The division between these regions marks the onset of a loss of QY (see section \ref{sec:QY_exp} and Fig.~\ref{fig:quantumyields_clustersize}).
Region I exhibits an average lifetime of about $32\,$ns for the lowest dopant density with a moderate monotonic lifetime reduction towards higher dopant density. 
In region II the lifetime decreases from approximately $25\,$ns at a surface dopant density of $0.02/\mathrm{nm}^2$ to approximately $2\,$ns at a density of $0.1/\mathrm{nm}^2$. On argon, the lifetime at low densities is approximately $23\,$ns. 
There is also a continuous decrease of the lifetimes, however, the lifetime reduction is not as strong as on neon. At a surface dopant density of $0.1/\mathrm{nm}^2$, where for neon the lifetime has decreased by more than one order of magnitude, on argon only a reduction by approximately $50\,\%$ is found. We note that the surface dopant density axis is on a logarithmic scale, following Ref.~\citenum{art:Izadnia17}.

Note that the cluster sizes used for argon and neon are comparable, as can be seen in the legend of Fig.~\ref{fig:lifetimes_clustersize}b).
This implies in particular that there is the same number of dopants and the same dopant density for the same cluster size.
For argon we could not obtain reliable lifetimes for larger dopant densities, because of low signal-to-noise ratios caused by the low fluorescence intensities.
Compared to neon, at low dopant densities, the lifetimes on argon are significantly reduced from $\approx32$\,ns on neon to $\approx23$\,ns on argon.
We also find that in contrast to neon, there is a dependence of the lifetimes on the cluster size; larger clusters have a longer lifetime.
We note, that one has to be careful in comparing the different data sets, since different laser powers have been used for the argon and the neon data. 
We will come back to this issue later in more detail.

\begin{figure}[tbp]
\centering
\includegraphics[width=8.2cm]{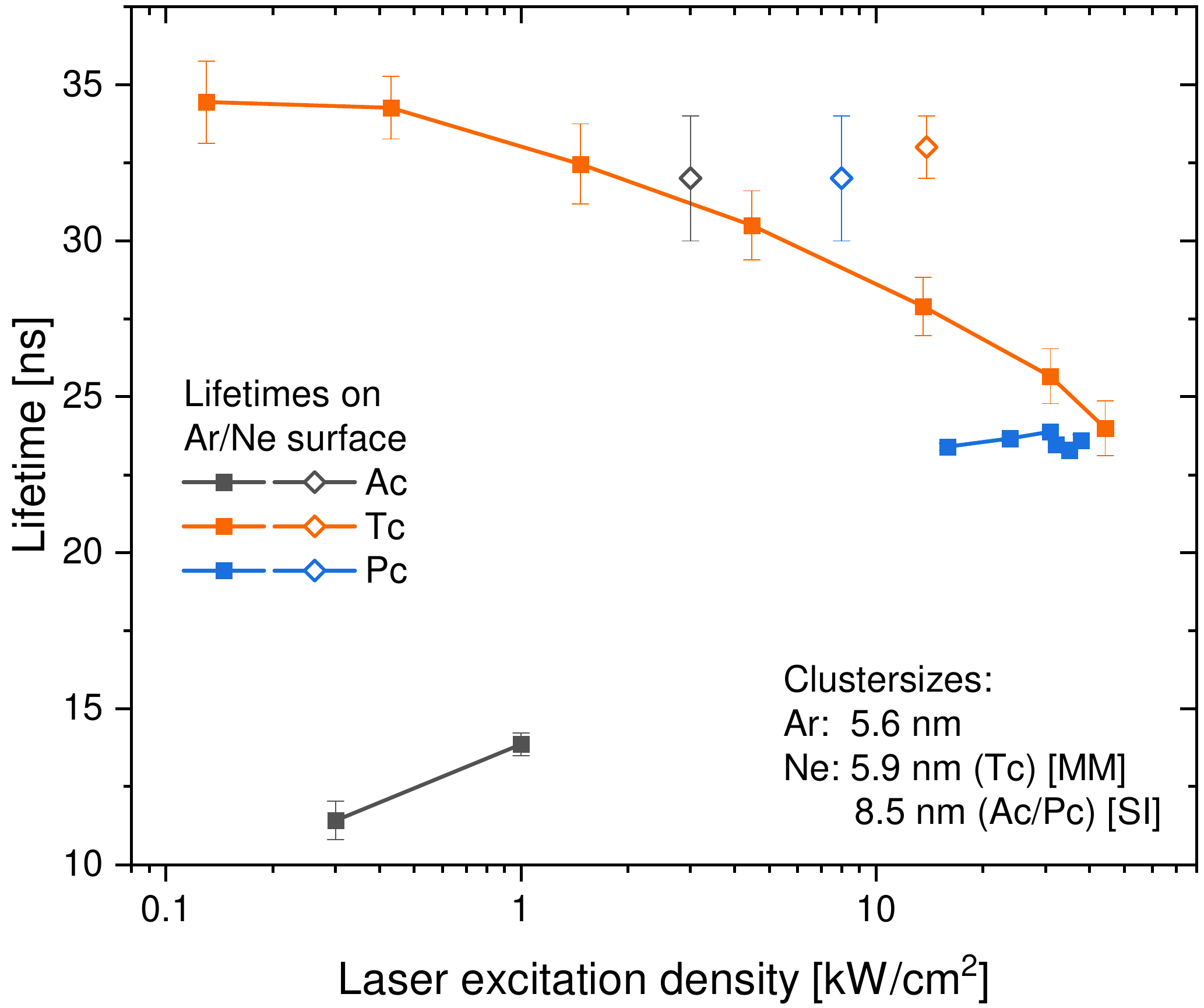}
\caption{Excitation lifetimes of Ac, Tc and Pc on Argon (filled squares) and Neon (empty diamonds) as a function of laser power density. Cluster sizes as indicated. Neon data, labeled [MM] and [SI] are taken from Refs.~\citenum{art:Mueller15} and \citenum{art:Izadnia17}, respectively.
For Tc on argon the measurements are for single molecules on the cluster (mean number of dopants per cluster of approximately $0.75$ or less).
Only for the the lowest laser power there are about 2 dopants at an average distance of about $8\,$nm. 
For Pc on argon there are on average approximately 1 dopant per cluster at an average distance of about $10\,$nm, except for the lowest laser power, where there are about 1.6 molecules on each cluster at an average distance of about $8\,$nm. For Ac, the average number of dopants is $1.5-2$, corresponding to average distances of $7-8\,$nm.
}
\label{fig:LifetimesVsLaserpower}
\end{figure}

\paragraph{Pentacene:}
The lifetimes of Pc show a similar behavior as the ones of Tc.
There is also a significant reduction of lifetime at low dopant densities when compared to Pc on neon. Similar to Tc also the lifetime reduction towards higher dopant densities is less pronounced than on neon. One notable difference to the Tc case is that for Pc the influence of the cluster size is significantly reduced.

\paragraph{Dependence on laser power:}
As already noted above, one has to be careful in comparing the different sets of data, since besides a variation in cluster size, also different laser powers have been used in different measurements. 
In Fig.~\ref{fig:LifetimesVsLaserpower} we show the dependence on the laser power in a regime of the dopant density where we expect that predominantly the molecules on the cluster remain isolated from each other. We show results where the cluster size is roughly the same for all combinations of Ac, Tc, and Pc as dopant molecules, and argon and neon as cluster materials. On neon all acenes have roughly the same lifetime in a narrow range of excitation densities. On argon there are huge differences of the lifetimes of the different acenes at the same laser power. For Tc on argon we performed a systematic scan of the laser power, which shows a rather strong decrease upon increasing the laser power.
Thus, we have to be careful comparing lifetime values of different measurements. However, the general shape of the lifetime functions with respect to the dopant density remains similar for different laser power density for both neon and argon data, respectively.
We would like to mention already here (see section \ref{sec:dis} for a detailed discussion) that the three acenes have very similar magnitudes of the transition dipole moment. Therefore, the number of excited molecules is directly proportional to the laser power.

Interestingly, the lifetime functions of Pc become virtually identical, when plotted with respect to the relative number of excited dopants (i.e., dopant density multiplied by laser power, not shown). 
A similar behavior is not found for Ac and Tc.

\subsection{Quantum yield of fluorescence}\label{sec:QY_exp}
\begin{figure}[tbp]
\centering
\includegraphics[width=8.2cm]{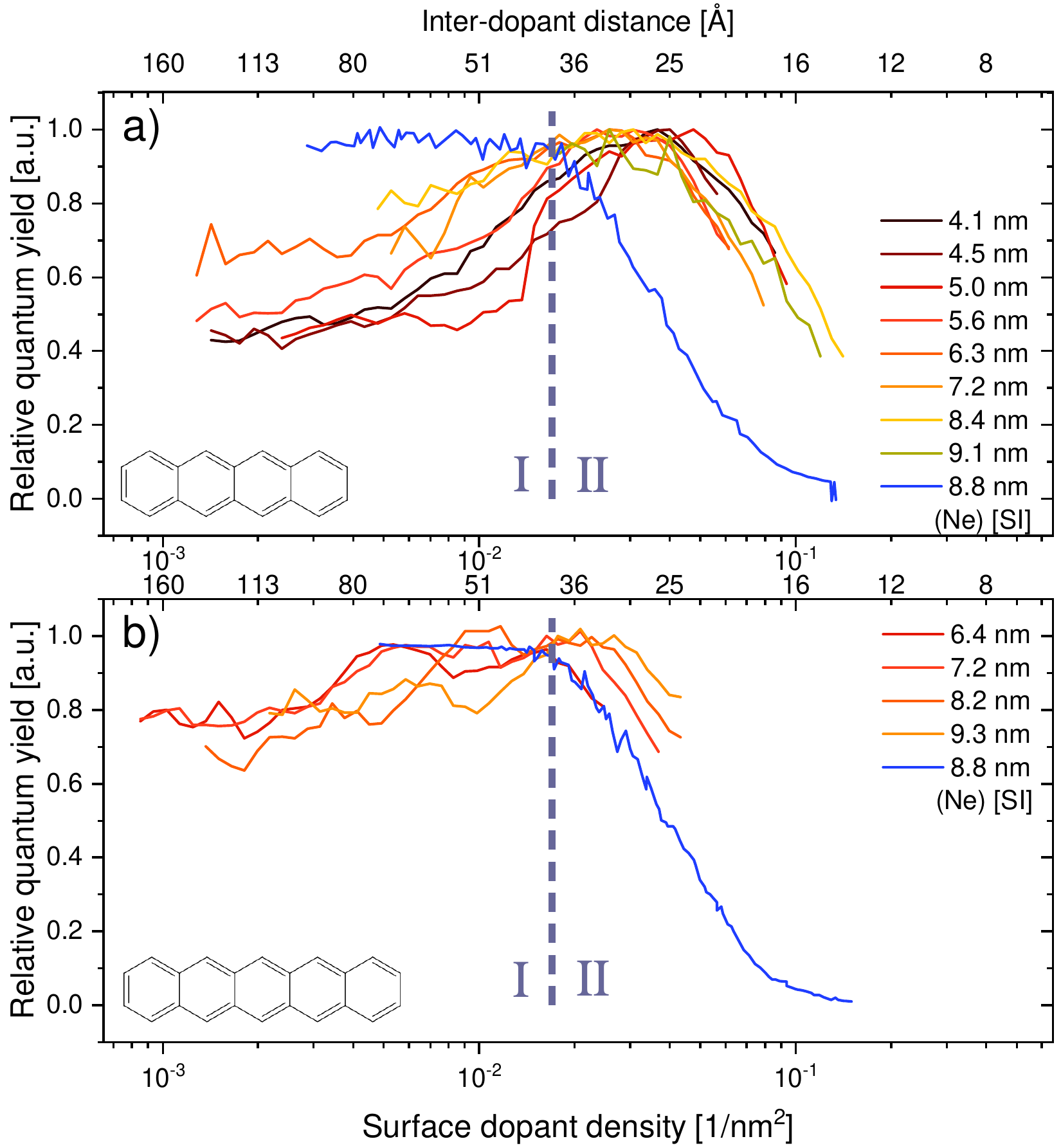}
\caption{Relative quantum yields as a function of the surface dopant density of a) Tc and b) Pc on argon and neon clusters of varying sizes. Neon data labeled [SI] are taken from Ref.~\citenum{art:Izadnia17}. The lower abscissas give the the surface dopant density, the upper abscissas give the respective inter-dopant distance.
}
\label{fig:quantumyields_clustersize}
\end{figure}

As mentioned in section~\ref{sec:met}, we are not able to measure the absolute QY. 
We are only able to measure the relative QY for different dopant densities at fixed cluster size. 
We consider in the following only Tc and Pc, because of a low signal-to-noise ratio in the case of Ac.
In Fig.~\ref{fig:quantumyields_clustersize} we present the relative QY as function of dopant density for different cluster sizes. 
We have normalized each data set to its respective maximum. 
One should keep in mind that therefore only the form of each function is significant, but the comparison of numerical values of different functions yields no meaningful information.

For orientation we have drawn the same vertical line as in Fig.~\ref{fig:lifetimes_clustersize} and distinguish two regimes, labeled I and II.

On the neon clusters, Tc and Pc have a very similar behavior:
In regime I,
below a surface dopant density of $\approx2\cdot 10^{-2}/\,\mathrm{nm}^{2}$, the relative QY is constant, above this density, in regime II, the relative QY drops to zero upon increase of the dopant density by about one order of magnitude. In contrast to neon, for both Tc and Pc on argon, the QY exhibits a rather broad maximum around the vertical line separating the two regimes.
Then, towards higher dopant densities the QY decreases. This decrease is similar to that on neon but starts at a larger dopant density (in particular for Tc). Also towards lower dopant densities the QY decreases. Below a surface dopant density of about $3 \cdot10^{-3}/$nm$^{2}$ the QY converges to a limiting value. Remarkably, this limiting value is significantly depending on the cluster size for Tc, showing a systematically lower QY for smaller clusters. For Pc, the limiting value of the QY does not depend on the cluster size.

To get a better feeling for the dopant density dependent number of molecules on the cluster, we remark that upon increase of the surface dopant density from $10^{-3}/\mathrm{nm^2}$ to $10^{-2}/\mathrm{nm^2}$, the mean number of dopants on the cluster increases from $0.5$ to $5$ for a cluster radius of approximately $6.3\,$nm. This means that at the lowest surface dopant density of $10^{-3}/\mathrm{nm^2}$, the signal can be safely assumed to be dominated by individual molecules.

Let us emphasize that Tc and Pc show on neon essentially the same relative quantum yields as a function of surface dopant density.
However, on argon only the general trends are similar, but e.g., for Pc the loss of QY sets in much earlier than for Tc.

\section{Discussion}\label{sec:dis}

\begin{figure}[tbp]
\centering
\includegraphics[width=8.2cm]{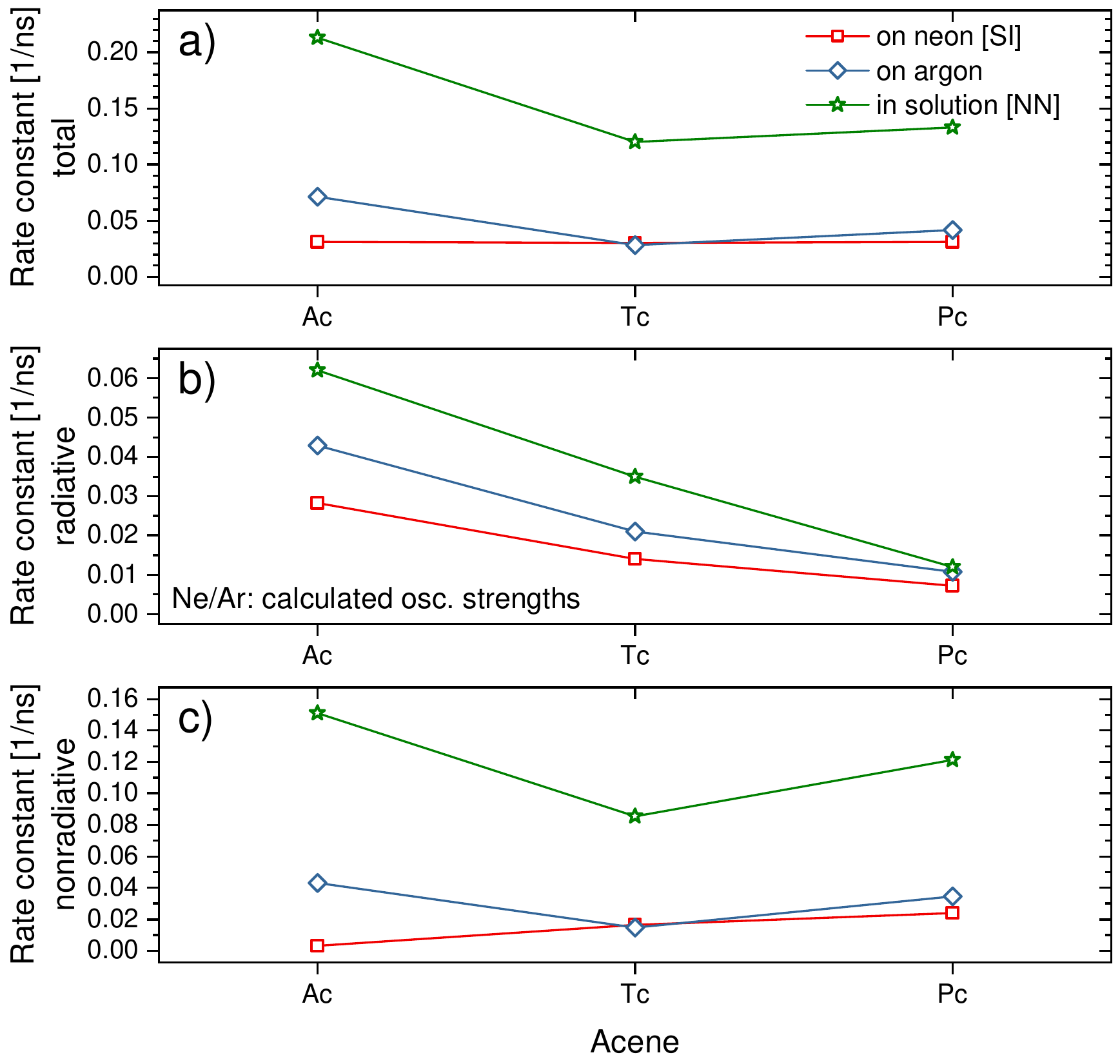}
\caption{Total (a), radiative (b) and non-radiative rate constants (c) of Ac, Tc and Pc in three different environments. 
For argon and neon the total rate constants are extracted from measurements (present work and Ref.~\citenum{art:Izadnia17}, respectively).
The radiative rate constants are numerically calculated as described in the text (see Eq.~(\ref{eqn:einstein-coeff})).
The non-radiative rate constants are deduced from Eq.~(\ref{eqn:totalrate}).
The rate constants of molecules in solution from Ref.~\citenum{art:Nijegorodov97} are deduced entirely from experimental results.
The uncertainty levels for the measured (total) lifetimes/rate constants are given to be less than $5\,\%$ for the solution data,\cite{art:Nijegorodov97} for the neon data\cite{art:Izadnia17} the lifetime fluctuations are below $10\,\%$, and for argon the lifetime fluctuations are below $15\,\%$. For the radiative and non-radiative rates in solution the uncertainties are estimated to be less than $15\,\%$.\cite{art:Nijegorodov97}}
\label{fig:rates_triple}
\end{figure}

Before we discuss our observations, we will give a brief overview of several mechanisms that might occur when several molecules are on a cluster. A particularly important interaction is the long-range Coulomb interaction between different molecules. This interaction can lead to the formation of collective states, where electronic excitation is coherently shared by several molecules.
For example, for two molecules such collective states are $|\pm\rangle
= (|S_1\rangle|S_0\rangle \pm |S_0\rangle|S_1\rangle)/\sqrt{2} $ and the corresponding transition strengths are proportional to $\vec{\mu}_1\pm\vec{\mu}_2$ with $\vec{\mu}_n$ being the transition dipole of molecule $n$.
Thus, these states can be sub- or superradiant depending on the orientation of the molecules with respect to each other.\cite{art:Eisfeld17} Furthermore, the energies of these collective states are shifted with respect to the energies of non-interacting molecules.
For large distances the Coulomb interaction takes the form of interacting dipoles.\cite{book:May2011}
The dipole-dipole interaction
depends on the arrangement of the molecules and scales as $\mu^2/R^3$, where $\mu$ denotes the magnitude of the transition dipole moment of the molecules and $R$ is the distance between molecules.
For all three acenes the transition dipoles connecting S$_0$ and S$_1$ are in the order of $1.9\,$Debye.
At distances around $1\,$nm the dipole-dipole-interaction ranges then from approximately $20$ to $-40$\,cm$^{-1}$, depending on the mutual arrangement. As the magnitude of this interaction energy is much smaller than the energy gap between the $0_0^0$ transition and the lowest vibrationally excited state, coupling to vibrational modes is strongly suppressed.
The interaction becomes stronger when molecules are closer together. For the molecules considered here, the interaction strength, based on the estimation of the dipole-dipole interaction strength of purely electronic states, becomes comparable to the energy gap between the vibrational ground state and the lowest vibrationally excited state (approximately $250\,\mathrm{cm}^{-1}$) in the electronic excited state at an inter-molecular distance of approximately $0.5\,$nm or below.
Therefore, at such small distances one expects changes of the absorption spectrum of the molecular assembly.
All effects caused by the coherent delocalization of electronic excitations are strongly diminished when the temperature becomes comparable to the interaction strength.\cite{art:Eisfeld17} Therefore, such collective effects will be more pronounced on the neon clusters. 
Even when coherence is not relevant, the interaction can still lead to an incoherent hopping of the excitation from one molecule to the other.
One should keep in mind that the substrate can slightly modify the strength of the interactions.
We address the importance of the dipole-dipole interaction below, in particular in connection with the LIF spectra.

Note, that we do not have direct information about the arrangement of the molecules. As the doping process is a statistical process, the positions and the orientations of the dopant molecules follow a statistical distribution. Individual molecules are assumed to be immobile, because of the low temperature of the clusters.
Whenever there are two or more dopant molecules on a cluster, there is a chance of two (or more) molecules being placed in close vicinity.
When this happens one expects strong changes in the absorption and emission spectra, as discussed above.
Furthermore,
interactions between close-by molecules could lead to a geometrical rearrangement of dopant molecules that are considered immobile at longer distances from each other.
Due to the interaction of molecules in their electronic excited state with other ground state molecules in close vicinity (depending on distances and orientation), a geometrical re-arrangement of the molecules can lead to the formation of an excimer state, which is energetically lower than the individual molecule's electronic excited state.

Individual molecules have, beside the radiative decay from the $S_1$ to the $S_0$ state, also non-radiative decay channels, e.g., {\it inter system crossing} (ISC) and {\it internal conversion} (IC).
Particularly, the interaction of acene molecules with argon atoms has been reported to significantly influence the rate constant of ISC and therefore also the sum of all non-radiative decay rate constants.\cite{art:Amirav86}
Also the interaction between molecules might change the corresponding rate constants.\cite{art:Celestino17}
Furthermore, several additional non-radiative de-excitation mechanisms are possible.
These are in particular {\it exciton-exciton annihilation} (EEA) and {\it singlet fission} (SF).
Also between collective states with multiple excitations, a change of transition strengths may occur (akin to the so-called Dicke superradiance\cite{art:Gross82}), which is similar to the sub- and superradiance of single-exciton states discussed above.
The previous results on neon clusters \cite{art:Izadnia17,art:Mueller15} have been interpreted in terms of an interplay between superradiance, exciton diffusion, EEA and SF, assuming that non-radiative decay channels of the individual molecules play a negligible role. 

In the following, we will discuss our experimental findings with respect to the mechanisms mentioned above including now also non-radiative decay chanels of the individual molecules.
We restrict our discussion to qualitative aspects, since a full solution of the many-body problem is out of hand and already reliable calculations for a single molecule attached to the clusters are difficult to achieve.

As became apparent from the discussion above, it is important to consider radiative and non-radiative contributions to the decay rate separately.
In this context, one should keep in mind, that although the time evolution of the fluorescence is recorded, the extracted time constant from the mono-exponential fit model is the inverse of the {\it total} transition rate constant (not only the radiative), i.e., $\tau_\text{recorded} = \tau_\text{tot} = 1/k_\text{tot}$.
Often one assumes the following relation between the total decay rate constant $k_\mathrm{tot}$ and the radiative and non-radiative rate constants:
\begin{equation}
\label{eqn:totalrate}
k_\text{tot} = k_\text{r} + k_\text{nr}
\end{equation}
In the following, a particular effort will be the separation of the radiative and the non-radiative decay constants.

\subsection{Lifetimes in the low doping limit -- extracting non-radiative decay constants for single molecules} \label{sec:low_Doping}
Before we consider the very complicated situation of several molecules on a cluster, we focus on the case of a single molecule on a cluster.
Our goal is to obtain rough estimates of the strength of radiative and non-radiative transition rates.
To this end we use Eq.~(\ref{eqn:totalrate}) with the total rate $k_\mathrm{tot}$ calculated as the inverse of the experimental lifetimes $\tau_\mathrm{recorded}$ and the radiative rate constant $k_\mathrm{r}$ is estimated from theoretical calculations.
Using these values, the non-radiative rate is obtained from $k_\mathrm{nr}=k_\mathrm{tot}-k_\mathrm{r}$.
We now discuss the extraction of $k_\mathrm{tot}$ from the experimentally obtained lifetimes. From Fig.~\ref{fig:number_vs_density_radius} we see that for clusters with radii smaller than 6.5\,nm and surface dopant densities smaller than $10^{-3}/\mathrm{nm}^{2}$, the probability to have more than a single molecule is sufficiently small. From Fig.~\ref{fig:lifetimes_clustersize} one sees that in this regime on neon all three acenes have a lifetime around 30\,ns.
On argon the lifetimes are significantly reduced. 
When interpreting the lifetimes on argon, some caution has to be taken regarding the laser-power.
While the data for Ac have been recorded using low laser power densities of approximately $1\,\mathrm{kW/cm}^2$, the data of Tc and Pc were obtained at laser power densities in the order of tens of kW/cm$^2$ (cf. Fig.~\ref{fig:lifetimes_clustersize}).
From the laser-power dependence of Fig.~\ref{fig:LifetimesVsLaserpower} we see that at least for Tc there is a pronounced dependence of the lifetime on laser-power.
One sees that for Tc on argon at low laser-power we obtain lifetimes comparable to those of single molecules on neon. 
For the following analysis we take for Tc the lifetime at the lowest laser power.
These lifetimes are displayed in Fig.~\ref{fig:rates_triple}a together with the values on neon and for comparison also lifetimes obtained by Nijegorodov et al.\cite{art:Nijegorodov97} in solution. 
These rate constants are about four times larger than on the clusters, with a decrease when going from Ac to Tc.
   
We now come to the radiative rate constant $k_\mathrm{r}$. For the solution case, the radiative rate is derived entirely from experimental results, as given in Ref.~\citenum{art:Nijegorodov97}. 
For the neon and the argon environment we estimate the radiative rate constant using\cite{art:Hilborn82}
\begin{equation}
\label{eqn:einstein-coeff}
k_\text{r} = \frac{2\pi \nu^2 e^2 n^3 f}{\varepsilon_0 m_\text{e} c^3}.
\end{equation}
Here $c$ is the speed of light, $e$ the elementary electric charge, $m_\mathrm{e}$ denotes the electron mass, and $\epsilon_0$ is the vacuum permittivity.
The refractive indices are taken to be\cite{art:Schulze74} $n=1.11$ for solid Ne and $n=1.29$ for solid Ar, $\nu$ is taken as the frequency of the respective $0_0^0$ transition (see Tab.~\ref{tab:00positions}). 
The values of the oscillator strengths are obtained from TDDFT calculations as $f = 0.041$ for Ac, $f = 0.031$ for Tc, and $f = 0.023$ for Pc.

The obtained rates are displayed in Fig.~\ref{fig:rates_triple}b.
One sees that the lifetimes decrease roughly by a factor 2, when going from Ac to Tc and from Tc to Pc (the rates are $0.028/0.014/0.007\,$ns$^{-1}$ for neon, and $0.042/0.021/0.011\,$ns$^{-1}$ for argon). According to Eq.~(\ref{eqn:einstein-coeff}) the difference between neon and argon stems only from different refractive indices.
We note that Eq.~(\ref{eqn:einstein-coeff}) is derived under the assumption that the molecule is completely surrounded by the medium. As an alternative model, we also calculated the radiative lifetimes from the decay of a dipole in front of a dielectric sphere.\cite{art:Chew87}
The resulting radiative transition rates in this dipole model are lower than the ones calculated from Eq.~(\ref{eqn:einstein-coeff}).
Assuming that the molecules are lying flat on the cluster, we found a decrease of about two thirds on argon and around one third on neon with respect to the values calculated from Eq.~(\ref{eqn:einstein-coeff}).
Under this assumption, the radiative rates for argon become even smaller than those for neon.
However, the systematic trend with respect to the length of the acenes remains similar.

Finally, the non-radiative rates calculated from $k_\mathrm{nr}=k_\mathrm{tot}-k_\mathrm{r}$ are shown in panel (c) of Fig.~\ref{fig:rates_triple}. For the solution data the non-radiative rates are about three times larger than the radiative rates, whereas for argon and neon they are comparable (for Ac on neon we even find that the radiative rate dominates).
For the case of neon clusters, the non-radiative rates increase approximately linearly with the length of the acenes.
In contrast, for argon and in solution we find a decrease when going from Ac to Tc followed by an increase when going from Tc to Pc.

\footnotetext[2]{In Ref.~\citenum{art:Nijegorodov97}, it was found that the rate constant for IC of Ac is more than one order of magnitude smaller compared to Tc and more than two orders of magnitude smaller compared to Pc, where it is the dominant decay channel}

To understand the behavior of the non-radiative rate constants, it would be desirable to further discriminate different non-radiative decay channels, in particular ISC and IC.
Unfortunately, for our measurements on neon and argon clusters, it is not possible to distinguish IC and ISC contributions, as it was done by Nijegorodov et al.\cite{art:Nijegorodov97} for the solution data.
There it was concluded that the ISC rate is decreasing from Ac to Pc and the IC rate is increasing in such a way that the overall non-radiative rate constant decreases from Ac to Tc and increases again from Tc to Pc.\cite{Note2}
The argon data resemble this behavior, where the neon data do not, indicating that argon with its higher dielectric constant might have the same trends as found in solution.

In particular for Ac, we see a strong enhancement of the non-radiative decay rate when changing the cluster material from neon to argon.
The reason for this strong change could be caused by shifts of triplet states relative to the S$_1$ state, which are caused by the increased environment interaction. 
A triplet state was estimated to be in the range of about $200\,$cm$^{-1}$ around the first excited singlet state.\cite{art:Widman72, art:Amirav86} 
A thermal activation of $200\,$cm$^{-1}$, as proposed by Widman and Huber,\cite{art:Widman72} is not possible in our experiment as the cluster temperatures are, as mentioned above, $10\,$K and $37\,$K for neon and argon, respectively.\cite{art:Farges81} 
As the shift of the 0-0 transition is increased upon change from neon to argon (see Tab.~\ref{tab:00positions}), it seems reasonable that also the triplet energies are significantly shifting, which could lead to an additional triplet state being energetically accessible.
 
The results at low dopant density, where there are only single molecules on the clusters, already highlight the importance of the material of the clusters. We see that it has an effect on both the radiative and the non-radiative transition rates.
When investigating the lifetimes as function of the dopant number, it is important to keep in mind that already individual molecules have on the cluster comparable or larger non-radiate decay rates than the radiative ones.
This is particularly pronounced for the argon clusters, but also present for neon.

We emphasize that the given values are supposed only to provide a rough estimation about the relative strength of the different acenes in the different environments. 
Despite that, important qualitative aspects, such as the large difference of the non-radiative rates for Ac on different clusters, are not affected by these uncertainties.

\subsection{LIF spectra -- formation of strongly interacting aggregates} \label{sec:Dis_LIF}
The interaction between molecules can alter the absorption spectra.
For example, for two molecules interacting via dipole-dipole interaction with strength $V$, one expects shifts of the $0_0^0$ line in the order of $V$, as long as $V$ is much smaller than the vibrational frequencies.
When the interaction energy becomes comparable to the vibrational spacing, significant changes of the spectra are expected.\cite{art:Roden11}
In the LIF spectra of Fig.~\ref{fig:spectra}, even at the highest dopant density the mean inter-molecule distance is on the order of $2.5\,\mathrm{nm}$.
At such mean distances the interaction is on the order of a few wavenumbers, which is much smaller than the vibrational frequencies, and it is even smaller when compared to the width of the peaks of the experimental spectra.
This is consistent with the observation that the lineshape of the $0_0^0$ transition is independent of the dopant density.
Also at higher frequencies all spectra except for Ac on argon are independent of the number of dopants and the surface dopant density.
This implies that the corresponding emission is either supressed or outside our detection range, if changes of the absorption spectra are present from strong interaction of molecules.
Ac on argon shows an exceptional behavior with a clear dependence on dopant density, suggesting that in this case the excited state and its dynamics is of different nature. 
Unfortunately, the spectra for Ac on neon from Ref.~\citenum{diss:Izadnia2018} do not cover the spectral region where the dopant-dependent changes appear in the argon system. 
Therefore, it is not clear, if the spectral changes are caused by the rare-gas interaction or if these appear from intrinsic properties of the acene molecules.

\subsection{Dopant density dependence of the radiative and non-radiative channels of Tc and Pc}
Now we turn our attention to multiply-doped clusters.
As discussed above, there are several dopant density-dependent mechanisms that can change the radiative and the non-radiative rate constants.
In the following, we discuss the dependence of these rates on the surface dopant density.

We first focus on the regime I of Figs.~\ref{fig:lifetimes_clustersize} and \ref{fig:quantumyields_clustersize}.
In this regime, for neon one has a constant QY and a strong decrease of total lifetime upon increasing the surface dopant density.
For argon, there is also a decrease in lifetime but an increase of the QY.
Such an increase of QY is possible, because the single molecules have a strong non-radiative decay and therefore a QY of less than 1.
In our previous studies\cite{art:Mueller15,art:Izadnia17} for neon, the lifetime reduction at constant QY was explained by an increase of superradiance with increasing surface dopant density.
In order to understand the increase of the QY due to superradiance, note that the QY can be written as
QY=$1-\left(k_\text{nr}/k_\text{tot}\right)=1-\left(k_\text{nr}/(k_\text{r}+k_\text{nr})\right)$. Provided a non-radiative rate independent of the dopant density, an increase of the radiative rate leads to an increase of the QY.
In section~\ref{sec:low_Doping} it can be seen that individual Tc and Pc molecules have non-radiative rate constants comparable to the radiative ones on both neon and argon.
Therefore, we would expect a comparable behavior of lifetime and QY for both molecules on neon and argon. Thus it is noteworthy that the data sets on neon and argon show pronounced differences. 
It is particularly surprising that Tc has a constant QY on neon but the QY is increasing on argon for increasing dopant density.
However, one has to keep in mind that calculated radiative rate constants are subject to a rather large uncertainty (as discussed in section \ref{sec:low_Doping}).
Since the change in QY depends sensitively on the exact numbers of $k_\text{r}$ and $k_\text{nr}$, the differences between Tc and Pc are consistent with the general conclusions of section \ref{sec:low_Doping} and consistent with the assumption that superradiance is appearing in region~I. 
Note, that upon increasing the dopant density, also the non-radiative decay constant changes.
Mechanisms such as SF and EEA increase the non-radiative decay rate and thus tend to decrease the QY (both mechanisms depend on the surface dopant density; EEA has a strong dependence on laser power as at least two excitations per cluster are needed).
In the border region between I and II, finally the non-radiative processes start dominating.
Again there are pronounced differences between argon and neon.
While for neon clusters the lifetime as function of the dopant density depends only weakly on the cluster size, for argon clusters the dependence is much more pronounced. Consequently, the dopant density at which the onset of the decay of the QY is found is very similar for neon clusters of different sizes and also the lifetimes corresponding to these dopant densities are very similar for neon. However, on argon the lifetimes exhibit a strong dependence on the cluster size. For small argon clusters, at the dopant density at which the decay of the QY sets in, the lifetime is significantly reduced. Contrastingly, for large argon clusters, the lifetime at that dopant density is still close to the lifetime of the low doping limit.
The lifetime in the case of argon exhibits a similar decrease for all cluster sizes, but is strongly shifted towards higher dopant densities and slightly longer lifetimes for larger clusters. The lifetime functions in the neon case are shifted similarly to the argon case, but to a much smaller extent.

We will now take a brief look at the influence of the cluster size on the excitation lifetimes and quantum yields.
From Fig.~\ref{fig:lifetimes_clustersize} one sees that for both Tc and Pc the lifetime is systematically higher at larger cluster sizes and the onset of decrease is shifted to larger dopant densities with increasing cluster size. In particular for Tc also the relative QY in region I is systematically higher for larger cluster sizes (cf.~Fig.~\ref{fig:quantumyields_clustersize}).
This indicates that with growing cluster size either the radiative decay is increased or the non-radiative decay is decreased.
At the border of region I and II, in the neon case the QY  starts to decrease with increasing dopant density. Opposed to that, in the argon case at this dopant density the QY exhibits a rather broad maximum. The onset of the decrease of the QY is at a higher dopant density, compared to the neon case. The decrease of the QY in region II is independent of the cluster size.

We note that for the argon case the behavior of the QY is independent of the cluster size in region II, but not in region I. This indicates that the decay dynamics is not solely depending on the dopant density, but also on the absolute number of dopants.
We do not dare to draw conclusions about the underlying mechanisms responsible for these differences.
However, the experimental results and the analysis presented above indicate that they are caused by the change of the material and the temperature of the cluster.

\subsection{The dopant density dependence of the lifetimes of Ac}\label{sec:Ac_diss}

We will now discuss the fundamentally different behavior of Ac on argon compared to neon. 
In particular, we address the {\it increase} of the lifetime on argon upon increasing the dopant density, in contrast to a {\it decrease} for neon.
As discussed in section \ref{sec:low_Doping}, we find for single Ac molecules on argon a much stronger non-radiative decay than for neon, which could be caused by a shift of triplet states relative to the $\mathrm{S}_1$ state. 
Note that, as mentioned in sec.~\ref{sec:previous}, SF is energetically forbidden for Ac and is therefore not available as a non-radiative decay channel.
Concerning the lifetime increase upon higher doping, one possible explanation is that the mutual interaction between the molecules induces further shifts of the triplet levels.
Another possible explanation is the formation of excimers, which has been discussed for amorphous Ac \cite{art:Sugino13, art:Ishii94} or in solution.\cite{art:Hofmann79} 
Hofmann et al.\cite{art:Hofmann79} reported on three distinct lifetime components of Ac excimer populations in solution with decay time constants of $10\,$ns, $35\,$ns, and $170\,$ns. 
The lifetime increase could then be explained by the creation of an excimer state.

The formation of excimers typically consists of two subsequent processes: the excitation of one individual molecule into its $S_1$ excited state as the first step, and subsequently the transformation of the individual molecules' $S_0$ and $S_1$ states into one joint excimer state.
As the actual excimer formation happens after the excitation, the absorption spectrum is not altered.
However, as has been discussed in section \ref{sec:Dis_LIF}, there is a pronounced change of the LIF spectra upon increasing the dopant density (cf.~Fig.~\ref{fig:spectra}a).
This spectral change could be caused by the transition dipole - dipole interaction leading to coherently delocalized excitations.
This also offers a possible mechanism for the increase in lifetime: the absorption process populates bright states, which subsequently relax to sub-radiant states.
Such sub-radiant states have been observed in theoretical studies of disordered particles.\cite{art:Abumwis20,art:Abumwis20a}
Both processes, the excimer formation and the formation of coherently delocalized states lead to a change of both radiative and non-radiative rate constants.
Such changes can appear, because for both processes there is a change in the accessible non-radiative channels caused by the interaction.
However, it remains surprising that these effects lead to opposite trends on argon and neon.

\section{Conclusion}\label{sec:concl}
In this work we have presented new insights in the interaction of polyacenes in a rare gas environment. 
The interaction of the chromophores with the environment significantly changes the excitation dynamics, although neon and argon are both considered as weakly interacting environments. 

Particularly, on argon the lifetime of Ac in the low doping limit is strongly reduced. It is approximately half of the lifetime as on neon. 
Also towards larger dopant densities the excitation dynamics differs strongly: on neon the lifetime {\it decreases} with  {\it increasing} dopant density, whereas it {\it increases} on argon.
Our findings imply that the previous model from Ref.~\citenum{art:Izadnia17} (including superradiance, singlet fission, exciton-exciton-annihilation and singlet hopping) in its present form is not sufficient to qualitatively explain the experimental data adequately. Non-radiative de-excitation channels of the individual molecules should be included in the model and the cluster material should be taken into account.
For Tc and Pc, for which SF is energetically allowed, an identical behavior on neon was found regarding excitation lifetimes and QY.\cite{art:Izadnia17} On argon, both molecules show quite distinct features regarding the cluster size dependence and the extent of the quenching channel at the lowest dopant densities, the excitation density dependence of the lifetime and the dependence of the lifetime on the number of excited dopants.

Furthermore, we discussed additional mechanisms including the population of an excimer state and the formation of strongly interacting aggregates in the context of the interpretation of the different excitation dynamics of a dense Ac ensemble compared to the dense ensembles of Tc or Pc. 
In future studies we would like to apply dispersed fluorescence spectroscopy to obtain information on possible excimer or aggregate species, and a pump probe scheme for the spectroscopy of triplet transitions, which could yield further information on ISC in Tc and Pc, in particular, with respect to the dopant density. 
Finally, REMPI (resonance enhanced multi-photon ionization) scheme experiments could give further information on the ratios of singlet and triplet populations.

\begin{acknowledgments}
\noindent 
The authors thank Johannes Fiedler and Stefan Y.~Buhmann for fruitful discussions and theoretical calculations.
This research was funded by the Deutsche Forschungsgemeinschaft (DFG) project STI 125/25-1 and the DFG-funded International Research Training Group CoCo (RTG 2079). M.~B.~gratefully acknowledges financial support by the Private Stiftung Ewald Marquardt. A.~E.~acknowledges support from the DFG via a Heisenberg fellowship (Grant No EI 872/5-1).
\end{acknowledgments}

\section*{Author Declarations}\label{sec:declaration}
\noindent
The authors have no conflicts to disclose.

\section*{Data availability statement}
\noindent
The data that support the findings of this study are available from the corresponding author upon reasonable request.


%
%

%


\bibliographystyle{aipnum4-1}

\bibliography{mybib}

\begin{thebibliography}{77}%
\makeatletter
\providecommand \@ifxundefined [1]{%
 \@ifx{#1\undefined}
}%
\providecommand \@ifnum [1]{%
 \ifnum #1\expandafter \@firstoftwo
 \else \expandafter \@secondoftwo
 \fi
}%
\providecommand \@ifx [1]{%
 \ifx #1\expandafter \@firstoftwo
 \else \expandafter \@secondoftwo
 \fi
}%
\providecommand \natexlab [1]{#1}%
\providecommand \enquote  [1]{``#1''}%
\providecommand \bibnamefont  [1]{#1}%
\providecommand \bibfnamefont [1]{#1}%
\providecommand \citenamefont [1]{#1}%
\providecommand \href@noop [0]{\@secondoftwo}%
\providecommand \href [0]{\begingroup \@sanitize@url \@href}%
\providecommand \@href[1]{\@@startlink{#1}\@@href}%
\providecommand \@@href[1]{\endgroup#1\@@endlink}%
\providecommand \@sanitize@url [0]{\catcode `\\12\catcode `\$12\catcode
  `\&12\catcode `\#12\catcode `\^12\catcode `\_12\catcode `\%12\relax}%
\providecommand \@@startlink[1]{}%
\providecommand \@@endlink[0]{}%
\providecommand \url  [0]{\begingroup\@sanitize@url \@url }%
\providecommand \@url [1]{\endgroup\@href {#1}{\urlprefix }}%
\providecommand \urlprefix  [0]{URL }%
\providecommand \Eprint [0]{\href }%
\providecommand \doibase [0]{http://dx.doi.org/}%
\providecommand \selectlanguage [0]{\@gobble}%
\providecommand \bibinfo  [0]{\@secondoftwo}%
\providecommand \bibfield  [0]{\@secondoftwo}%
\providecommand \translation [1]{[#1]}%
\providecommand \BibitemOpen [0]{}%
\providecommand \bibitemStop [0]{}%
\providecommand \bibitemNoStop [0]{.\EOS\space}%
\providecommand \EOS [0]{\spacefactor3000\relax}%
\providecommand \BibitemShut  [1]{\csname bibitem#1\endcsname}%
\let\auto@bib@innerbib\@empty
\bibitem [{\citenamefont {Nelson}\ \emph {et~al.}(1998)\citenamefont {Nelson},
  \citenamefont {Lin}, \citenamefont {Gundlach},\ and\ \citenamefont
  {Jackson}}]{art:Nelson98}%
  \BibitemOpen
  \bibfield  {author} {\bibinfo {author} {\bibfnamefont {S.~F.}\ \bibnamefont
  {Nelson}}, \bibinfo {author} {\bibfnamefont {Y.-Y.}\ \bibnamefont {Lin}},
  \bibinfo {author} {\bibfnamefont {D.~J.}\ \bibnamefont {Gundlach}}, \ and\
  \bibinfo {author} {\bibfnamefont {T.~N.}\ \bibnamefont {Jackson}},\ }\href
  {\doibase 10.1063/1.121205} {\bibfield  {journal} {\bibinfo  {journal}
  {Applied Physics Letters}\ }\textbf {\bibinfo {volume} {72}},\ \bibinfo
  {pages} {1854} (\bibinfo {year} {1998})}\BibitemShut {NoStop}%
\bibitem [{\citenamefont {Li}\ \emph {et~al.}(2012)\citenamefont {Li},
  \citenamefont {Wang}, \citenamefont {Ren}, \citenamefont {Gao}, \citenamefont
  {Hong}, \citenamefont {Li},\ and\ \citenamefont {Zhu}}]{art:Li12}%
  \BibitemOpen
  \bibfield  {author} {\bibinfo {author} {\bibfnamefont {J.}~\bibnamefont
  {Li}}, \bibinfo {author} {\bibfnamefont {M.}~\bibnamefont {Wang}}, \bibinfo
  {author} {\bibfnamefont {S.}~\bibnamefont {Ren}}, \bibinfo {author}
  {\bibfnamefont {X.}~\bibnamefont {Gao}}, \bibinfo {author} {\bibfnamefont
  {W.}~\bibnamefont {Hong}}, \bibinfo {author} {\bibfnamefont {H.}~\bibnamefont
  {Li}}, \ and\ \bibinfo {author} {\bibfnamefont {D.}~\bibnamefont {Zhu}},\
  }\href {\doibase 10.1039/c2jm16871e} {\bibfield  {journal} {\bibinfo
  {journal} {Journal of Materials Chemistry}\ }\textbf {\bibinfo {volume}
  {22}},\ \bibinfo {pages} {10496} (\bibinfo {year} {2012})}\BibitemShut
  {NoStop}%
\bibitem [{\citenamefont {Lee}\ \emph {et~al.}(2019)\citenamefont {Lee},
  \citenamefont {Lee}, \citenamefont {Lee}, \citenamefont {Oh}, \citenamefont
  {Choi}, \citenamefont {Im}, \citenamefont {Jang},\ and\ \citenamefont
  {Kim}}]{art:Lee19}%
  \BibitemOpen
  \bibfield  {author} {\bibinfo {author} {\bibfnamefont {K.~H.}\ \bibnamefont
  {Lee}}, \bibinfo {author} {\bibfnamefont {K.}~\bibnamefont {Lee}}, \bibinfo
  {author} {\bibfnamefont {G.}~\bibnamefont {Lee}}, \bibinfo {author}
  {\bibfnamefont {M.~S.}\ \bibnamefont {Oh}}, \bibinfo {author} {\bibfnamefont
  {J.-M.}\ \bibnamefont {Choi}}, \bibinfo {author} {\bibfnamefont
  {S.}~\bibnamefont {Im}}, \bibinfo {author} {\bibfnamefont {S.}~\bibnamefont
  {Jang}}, \ and\ \bibinfo {author} {\bibfnamefont {E.}~\bibnamefont {Kim}},\
  }\href {\doibase 10.1149/1.2980558} {\bibfield  {journal} {\bibinfo
  {journal} {{ECS} Transactions}\ }\textbf {\bibinfo {volume} {16}},\ \bibinfo
  {pages} {239} (\bibinfo {year} {2019})}\BibitemShut {NoStop}%
\bibitem [{\citenamefont {Xia}\ \emph {et~al.}(2016)\citenamefont {Xia},
  \citenamefont {Sanders}, \citenamefont {Cheng}, \citenamefont {Low},
  \citenamefont {Liu}, \citenamefont {Campos},\ and\ \citenamefont
  {Sun}}]{art:Xia16}%
  \BibitemOpen
  \bibfield  {author} {\bibinfo {author} {\bibfnamefont {J.}~\bibnamefont
  {Xia}}, \bibinfo {author} {\bibfnamefont {S.~N.}\ \bibnamefont {Sanders}},
  \bibinfo {author} {\bibfnamefont {W.}~\bibnamefont {Cheng}}, \bibinfo
  {author} {\bibfnamefont {J.~Z.}\ \bibnamefont {Low}}, \bibinfo {author}
  {\bibfnamefont {J.}~\bibnamefont {Liu}}, \bibinfo {author} {\bibfnamefont
  {L.~M.}\ \bibnamefont {Campos}}, \ and\ \bibinfo {author} {\bibfnamefont
  {T.}~\bibnamefont {Sun}},\ }\href {\doibase 10.1002/adma.201601652}
  {\bibfield  {journal} {\bibinfo  {journal} {Advanced Materials}\ }\textbf
  {\bibinfo {volume} {29}},\ \bibinfo {pages} {1601652} (\bibinfo {year}
  {2016})}\BibitemShut {NoStop}%
\bibitem [{\citenamefont {Thompson}\ \emph {et~al.}(2013)\citenamefont
  {Thompson}, \citenamefont {Congreve}, \citenamefont {Goldberg}, \citenamefont
  {Menon},\ and\ \citenamefont {Baldo}}]{art:Thompson13}%
  \BibitemOpen
  \bibfield  {author} {\bibinfo {author} {\bibfnamefont {N.~J.}\ \bibnamefont
  {Thompson}}, \bibinfo {author} {\bibfnamefont {D.~N.}\ \bibnamefont
  {Congreve}}, \bibinfo {author} {\bibfnamefont {D.}~\bibnamefont {Goldberg}},
  \bibinfo {author} {\bibfnamefont {V.~M.}\ \bibnamefont {Menon}}, \ and\
  \bibinfo {author} {\bibfnamefont {M.~A.}\ \bibnamefont {Baldo}},\ }\href
  {\doibase 10.1063/1.4858176} {\bibfield  {journal} {\bibinfo  {journal}
  {Applied Physics Letters}\ }\textbf {\bibinfo {volume} {103}},\ \bibinfo
  {pages} {263302} (\bibinfo {year} {2013})}\BibitemShut {NoStop}%
\bibitem [{\citenamefont {Walker}\ \emph {et~al.}(2013)\citenamefont {Walker},
  \citenamefont {Musser}, \citenamefont {Beljonne},\ and\ \citenamefont
  {Friend}}]{art:Walker13}%
  \BibitemOpen
  \bibfield  {author} {\bibinfo {author} {\bibfnamefont {B.~J.}\ \bibnamefont
  {Walker}}, \bibinfo {author} {\bibfnamefont {A.~J.}\ \bibnamefont {Musser}},
  \bibinfo {author} {\bibfnamefont {D.}~\bibnamefont {Beljonne}}, \ and\
  \bibinfo {author} {\bibfnamefont {R.~H.}\ \bibnamefont {Friend}},\ }\href
  {\doibase 10.1038/nchem.1801} {\bibfield  {journal} {\bibinfo  {journal}
  {Nature Chemistry}\ }\textbf {\bibinfo {volume} {5}},\ \bibinfo {pages}
  {1019} (\bibinfo {year} {2013})}\BibitemShut {NoStop}%
\bibitem [{\citenamefont {Irimia-Vladu}\ \emph {et~al.}(2010)\citenamefont
  {Irimia-Vladu}, \citenamefont {Troshin}, \citenamefont {Reisinger},
  \citenamefont {Shmygleva}, \citenamefont {Kanbur}, \citenamefont
  {Schwabegger}, \citenamefont {Bodea}, \citenamefont {Schwödiauer},
  \citenamefont {Mumyatov}, \citenamefont {Fergus}, \citenamefont {Razumov},
  \citenamefont {Sitter}, \citenamefont {Sariciftci},\ and\ \citenamefont
  {Bauer}}]{art:Irimia-Vladu10}%
  \BibitemOpen
  \bibfield  {author} {\bibinfo {author} {\bibfnamefont {M.}~\bibnamefont
  {Irimia-Vladu}}, \bibinfo {author} {\bibfnamefont {P.~A.}\ \bibnamefont
  {Troshin}}, \bibinfo {author} {\bibfnamefont {M.}~\bibnamefont {Reisinger}},
  \bibinfo {author} {\bibfnamefont {L.}~\bibnamefont {Shmygleva}}, \bibinfo
  {author} {\bibfnamefont {Y.}~\bibnamefont {Kanbur}}, \bibinfo {author}
  {\bibfnamefont {G.}~\bibnamefont {Schwabegger}}, \bibinfo {author}
  {\bibfnamefont {M.}~\bibnamefont {Bodea}}, \bibinfo {author} {\bibfnamefont
  {R.}~\bibnamefont {Schwödiauer}}, \bibinfo {author} {\bibfnamefont
  {A.}~\bibnamefont {Mumyatov}}, \bibinfo {author} {\bibfnamefont {J.~W.}\
  \bibnamefont {Fergus}}, \bibinfo {author} {\bibfnamefont {V.~F.}\
  \bibnamefont {Razumov}}, \bibinfo {author} {\bibfnamefont {H.}~\bibnamefont
  {Sitter}}, \bibinfo {author} {\bibfnamefont {N.~S.}\ \bibnamefont
  {Sariciftci}}, \ and\ \bibinfo {author} {\bibfnamefont {S.}~\bibnamefont
  {Bauer}},\ }\href {\doibase 10.1002/adfm.201001031} {\bibfield  {journal}
  {\bibinfo  {journal} {Advanced Functional Materials}\ }\textbf {\bibinfo
  {volume} {20}},\ \bibinfo {pages} {4069} (\bibinfo {year}
  {2010})}\BibitemShut {NoStop}%
\bibitem [{\citenamefont {Zan}\ \emph {et~al.}(2012)\citenamefont {Zan},
  \citenamefont {Tsai}, \citenamefont {ren Lo}, \citenamefont {Wu},\ and\
  \citenamefont {Yang}}]{art:Zan12}%
  \BibitemOpen
  \bibfield  {author} {\bibinfo {author} {\bibfnamefont {H.-W.}\ \bibnamefont
  {Zan}}, \bibinfo {author} {\bibfnamefont {W.-W.}\ \bibnamefont {Tsai}},
  \bibinfo {author} {\bibfnamefont {Y.}~\bibnamefont {ren Lo}}, \bibinfo
  {author} {\bibfnamefont {Y.-M.}\ \bibnamefont {Wu}}, \ and\ \bibinfo {author}
  {\bibfnamefont {Y.-S.}\ \bibnamefont {Yang}},\ }\href {\doibase
  10.1109/jsen.2011.2121901} {\bibfield  {journal} {\bibinfo  {journal} {{IEEE}
  Sensors Journal}\ }\textbf {\bibinfo {volume} {12}},\ \bibinfo {pages} {594}
  (\bibinfo {year} {2012})}\BibitemShut {NoStop}%
\bibitem [{\citenamefont {Baude}\ \emph {et~al.}(2003)\citenamefont {Baude},
  \citenamefont {Ender}, \citenamefont {Haase}, \citenamefont {Kelley},
  \citenamefont {Muyres},\ and\ \citenamefont {Theiss}}]{art:Baude03}%
  \BibitemOpen
  \bibfield  {author} {\bibinfo {author} {\bibfnamefont {P.~F.}\ \bibnamefont
  {Baude}}, \bibinfo {author} {\bibfnamefont {D.~A.}\ \bibnamefont {Ender}},
  \bibinfo {author} {\bibfnamefont {M.~A.}\ \bibnamefont {Haase}}, \bibinfo
  {author} {\bibfnamefont {T.~W.}\ \bibnamefont {Kelley}}, \bibinfo {author}
  {\bibfnamefont {D.~V.}\ \bibnamefont {Muyres}}, \ and\ \bibinfo {author}
  {\bibfnamefont {S.~D.}\ \bibnamefont {Theiss}},\ }\href {\doibase
  10.1063/1.1579554} {\bibfield  {journal} {\bibinfo  {journal} {Applied
  Physics Letters}\ }\textbf {\bibinfo {volume} {82}},\ \bibinfo {pages} {3964}
  (\bibinfo {year} {2003})}\BibitemShut {NoStop}%
\bibitem [{\citenamefont {Nomura}\ \emph {et~al.}(2004)\citenamefont {Nomura},
  \citenamefont {Ohta}, \citenamefont {Takagi}, \citenamefont {Kamiya},
  \citenamefont {Hirano},\ and\ \citenamefont {Hosono}}]{art:Nomura04}%
  \BibitemOpen
  \bibfield  {author} {\bibinfo {author} {\bibfnamefont {K.}~\bibnamefont
  {Nomura}}, \bibinfo {author} {\bibfnamefont {H.}~\bibnamefont {Ohta}},
  \bibinfo {author} {\bibfnamefont {A.}~\bibnamefont {Takagi}}, \bibinfo
  {author} {\bibfnamefont {T.}~\bibnamefont {Kamiya}}, \bibinfo {author}
  {\bibfnamefont {M.}~\bibnamefont {Hirano}}, \ and\ \bibinfo {author}
  {\bibfnamefont {H.}~\bibnamefont {Hosono}},\ }\href {\doibase
  10.1038/nature03090} {\bibfield  {journal} {\bibinfo  {journal} {Nature}\
  }\textbf {\bibinfo {volume} {432}},\ \bibinfo {pages} {488} (\bibinfo {year}
  {2004})}\BibitemShut {NoStop}%
\bibitem [{\citenamefont {Lassnig}\ \emph {et~al.}(2015)\citenamefont
  {Lassnig}, \citenamefont {Hollerer}, \citenamefont {Striedinger},
  \citenamefont {Fian}, \citenamefont {Stadlober},\ and\ \citenamefont
  {Winkler}}]{art:Lassnig15}%
  \BibitemOpen
  \bibfield  {author} {\bibinfo {author} {\bibfnamefont {R.}~\bibnamefont
  {Lassnig}}, \bibinfo {author} {\bibfnamefont {M.}~\bibnamefont {Hollerer}},
  \bibinfo {author} {\bibfnamefont {B.}~\bibnamefont {Striedinger}}, \bibinfo
  {author} {\bibfnamefont {A.}~\bibnamefont {Fian}}, \bibinfo {author}
  {\bibfnamefont {B.}~\bibnamefont {Stadlober}}, \ and\ \bibinfo {author}
  {\bibfnamefont {A.}~\bibnamefont {Winkler}},\ }\href {\doibase
  10.1016/j.orgel.2015.08.016} {\bibfield  {journal} {\bibinfo  {journal}
  {Organic Electronics}\ }\textbf {\bibinfo {volume} {26}},\ \bibinfo {pages}
  {420} (\bibinfo {year} {2015})}\BibitemShut {NoStop}%
\bibitem [{\citenamefont {Lin}\ \emph {et~al.}(1997{\natexlab{a}})\citenamefont
  {Lin}, \citenamefont {Gundlach}, \citenamefont {Nelson},\ and\ \citenamefont
  {Jackson}}]{art:Lin97}%
  \BibitemOpen
  \bibfield  {author} {\bibinfo {author} {\bibfnamefont {Y.-Y.}\ \bibnamefont
  {Lin}}, \bibinfo {author} {\bibfnamefont {D.}~\bibnamefont {Gundlach}},
  \bibinfo {author} {\bibfnamefont {S.}~\bibnamefont {Nelson}}, \ and\ \bibinfo
  {author} {\bibfnamefont {T.}~\bibnamefont {Jackson}},\ }\href {\doibase
  10.1109/16.605476} {\bibfield  {journal} {\bibinfo  {journal} {{IEEE}
  Transactions on Electron Devices}\ }\textbf {\bibinfo {volume} {44}},\
  \bibinfo {pages} {1325} (\bibinfo {year} {1997}{\natexlab{a}})}\BibitemShut
  {NoStop}%
\bibitem [{\citenamefont {Lin}\ \emph {et~al.}(1997{\natexlab{b}})\citenamefont
  {Lin}, \citenamefont {Gundlach}, \citenamefont {Nelson},\ and\ \citenamefont
  {Jackson}}]{art:Lin97a}%
  \BibitemOpen
  \bibfield  {author} {\bibinfo {author} {\bibfnamefont {Y.-Y.}\ \bibnamefont
  {Lin}}, \bibinfo {author} {\bibfnamefont {D.}~\bibnamefont {Gundlach}},
  \bibinfo {author} {\bibfnamefont {S.}~\bibnamefont {Nelson}}, \ and\ \bibinfo
  {author} {\bibfnamefont {T.}~\bibnamefont {Jackson}},\ }\href {\doibase
  10.1109/55.644085} {\bibfield  {journal} {\bibinfo  {journal} {{IEEE}
  Electron Device Letters}\ }\textbf {\bibinfo {volume} {18}},\ \bibinfo
  {pages} {606} (\bibinfo {year} {1997}{\natexlab{b}})}\BibitemShut {NoStop}%
\bibitem [{\citenamefont {Tan}\ \emph {et~al.}(2009)\citenamefont {Tan},
  \citenamefont {Mathews}, \citenamefont {Cahyadi}, \citenamefont {Zhu},\ and\
  \citenamefont {Mhaisalkar}}]{art:Tan09}%
  \BibitemOpen
  \bibfield  {author} {\bibinfo {author} {\bibfnamefont {H.~S.}\ \bibnamefont
  {Tan}}, \bibinfo {author} {\bibfnamefont {N.}~\bibnamefont {Mathews}},
  \bibinfo {author} {\bibfnamefont {T.}~\bibnamefont {Cahyadi}}, \bibinfo
  {author} {\bibfnamefont {F.~R.}\ \bibnamefont {Zhu}}, \ and\ \bibinfo
  {author} {\bibfnamefont {S.~G.}\ \bibnamefont {Mhaisalkar}},\ }\href
  {\doibase 10.1063/1.3168523} {\bibfield  {journal} {\bibinfo  {journal}
  {Applied Physics Letters}\ }\textbf {\bibinfo {volume} {94}},\ \bibinfo
  {pages} {263303} (\bibinfo {year} {2009})}\BibitemShut {NoStop}%
\bibitem [{\citenamefont {Khan}\ \emph {et~al.}(2011)\citenamefont {Khan},
  \citenamefont {Roberts}, \citenamefont {Knoll},\ and\ \citenamefont
  {Bao}}]{art:Khan11}%
  \BibitemOpen
  \bibfield  {author} {\bibinfo {author} {\bibfnamefont {H.~U.}\ \bibnamefont
  {Khan}}, \bibinfo {author} {\bibfnamefont {M.~E.}\ \bibnamefont {Roberts}},
  \bibinfo {author} {\bibfnamefont {W.}~\bibnamefont {Knoll}}, \ and\ \bibinfo
  {author} {\bibfnamefont {Z.}~\bibnamefont {Bao}},\ }\href {\doibase
  10.1021/cm103685c} {\bibfield  {journal} {\bibinfo  {journal} {Chemistry of
  Materials}\ }\textbf {\bibinfo {volume} {23}},\ \bibinfo {pages} {1946}
  (\bibinfo {year} {2011})}\BibitemShut {NoStop}%
\bibitem [{\citenamefont {Sheraw}\ \emph {et~al.}(2002)\citenamefont {Sheraw},
  \citenamefont {Zhou}, \citenamefont {Huang}, \citenamefont {Gundlach},
  \citenamefont {Jackson}, \citenamefont {Kane}, \citenamefont {Hill},
  \citenamefont {Hammond}, \citenamefont {Campi}, \citenamefont {Greening},
  \citenamefont {Francl},\ and\ \citenamefont {West}}]{art:Sheraw02}%
  \BibitemOpen
  \bibfield  {author} {\bibinfo {author} {\bibfnamefont {C.~D.}\ \bibnamefont
  {Sheraw}}, \bibinfo {author} {\bibfnamefont {L.}~\bibnamefont {Zhou}},
  \bibinfo {author} {\bibfnamefont {J.~R.}\ \bibnamefont {Huang}}, \bibinfo
  {author} {\bibfnamefont {D.~J.}\ \bibnamefont {Gundlach}}, \bibinfo {author}
  {\bibfnamefont {T.~N.}\ \bibnamefont {Jackson}}, \bibinfo {author}
  {\bibfnamefont {M.~G.}\ \bibnamefont {Kane}}, \bibinfo {author}
  {\bibfnamefont {I.~G.}\ \bibnamefont {Hill}}, \bibinfo {author}
  {\bibfnamefont {M.~S.}\ \bibnamefont {Hammond}}, \bibinfo {author}
  {\bibfnamefont {J.}~\bibnamefont {Campi}}, \bibinfo {author} {\bibfnamefont
  {B.~K.}\ \bibnamefont {Greening}}, \bibinfo {author} {\bibfnamefont
  {J.}~\bibnamefont {Francl}}, \ and\ \bibinfo {author} {\bibfnamefont
  {J.}~\bibnamefont {West}},\ }\href {\doibase 10.1063/1.1448659} {\bibfield
  {journal} {\bibinfo  {journal} {Applied Physics Letters}\ }\textbf {\bibinfo
  {volume} {80}},\ \bibinfo {pages} {1088} (\bibinfo {year}
  {2002})}\BibitemShut {NoStop}%
\bibitem [{\citenamefont {Rogers}\ \emph {et~al.}(2001)\citenamefont {Rogers},
  \citenamefont {Bao}, \citenamefont {Baldwin}, \citenamefont {Dodabalapur},
  \citenamefont {Crone}, \citenamefont {Raju}, \citenamefont {Kuck},
  \citenamefont {Katz}, \citenamefont {Amundson}, \citenamefont {Ewing},\ and\
  \citenamefont {Drzaic}}]{art:Rogers01}%
  \BibitemOpen
  \bibfield  {author} {\bibinfo {author} {\bibfnamefont {J.~A.}\ \bibnamefont
  {Rogers}}, \bibinfo {author} {\bibfnamefont {Z.}~\bibnamefont {Bao}},
  \bibinfo {author} {\bibfnamefont {K.}~\bibnamefont {Baldwin}}, \bibinfo
  {author} {\bibfnamefont {A.}~\bibnamefont {Dodabalapur}}, \bibinfo {author}
  {\bibfnamefont {B.}~\bibnamefont {Crone}}, \bibinfo {author} {\bibfnamefont
  {V.~R.}\ \bibnamefont {Raju}}, \bibinfo {author} {\bibfnamefont
  {V.}~\bibnamefont {Kuck}}, \bibinfo {author} {\bibfnamefont {H.}~\bibnamefont
  {Katz}}, \bibinfo {author} {\bibfnamefont {K.}~\bibnamefont {Amundson}},
  \bibinfo {author} {\bibfnamefont {J.}~\bibnamefont {Ewing}}, \ and\ \bibinfo
  {author} {\bibfnamefont {P.}~\bibnamefont {Drzaic}},\ }\href {\doibase
  10.1073/pnas.091588098} {\bibfield  {journal} {\bibinfo  {journal}
  {Proceedings of the National Academy of Sciences}\ }\textbf {\bibinfo
  {volume} {98}},\ \bibinfo {pages} {4835} (\bibinfo {year}
  {2001})}\BibitemShut {NoStop}%
\bibitem [{\citenamefont {Someya}\ \emph {et~al.}(2005)\citenamefont {Someya},
  \citenamefont {Kato}, \citenamefont {Sekitani}, \citenamefont {Iba},
  \citenamefont {Noguchi}, \citenamefont {Murase}, \citenamefont {Kawaguchi},\
  and\ \citenamefont {Sakurai}}]{art:Someya05}%
  \BibitemOpen
  \bibfield  {author} {\bibinfo {author} {\bibfnamefont {T.}~\bibnamefont
  {Someya}}, \bibinfo {author} {\bibfnamefont {Y.}~\bibnamefont {Kato}},
  \bibinfo {author} {\bibfnamefont {T.}~\bibnamefont {Sekitani}}, \bibinfo
  {author} {\bibfnamefont {S.}~\bibnamefont {Iba}}, \bibinfo {author}
  {\bibfnamefont {Y.}~\bibnamefont {Noguchi}}, \bibinfo {author} {\bibfnamefont
  {Y.}~\bibnamefont {Murase}}, \bibinfo {author} {\bibfnamefont
  {H.}~\bibnamefont {Kawaguchi}}, \ and\ \bibinfo {author} {\bibfnamefont
  {T.}~\bibnamefont {Sakurai}},\ }\href {\doibase 10.1073/pnas.0502392102}
  {\bibfield  {journal} {\bibinfo  {journal} {Proceedings of the National
  Academy of Sciences}\ }\textbf {\bibinfo {volume} {102}},\ \bibinfo {pages}
  {12321} (\bibinfo {year} {2005})}\BibitemShut {NoStop}%
\bibitem [{\citenamefont {Peter}\ and\ \citenamefont
  {Bässler}(1980)}]{art:Peter80}%
  \BibitemOpen
  \bibfield  {author} {\bibinfo {author} {\bibfnamefont {G.}~\bibnamefont
  {Peter}}\ and\ \bibinfo {author} {\bibfnamefont {H.}~\bibnamefont
  {Bässler}},\ }\href {\doibase 10.1016/0301-0104(80)85033-6} {\bibfield
  {journal} {\bibinfo  {journal} {Chemical Physics}\ }\textbf {\bibinfo
  {volume} {49}},\ \bibinfo {pages} {9} (\bibinfo {year} {1980})}\BibitemShut
  {NoStop}%
\bibitem [{\citenamefont {Amirav}, \citenamefont {Even},\ and\ \citenamefont
  {Jortner}(1982)}]{art:Amirav82}%
  \BibitemOpen
  \bibfield  {author} {\bibinfo {author} {\bibfnamefont {A.}~\bibnamefont
  {Amirav}}, \bibinfo {author} {\bibfnamefont {U.}~\bibnamefont {Even}}, \ and\
  \bibinfo {author} {\bibfnamefont {J.}~\bibnamefont {Jortner}},\ }\href
  {\doibase 10.1021/j100214a017} {\bibfield  {journal} {\bibinfo  {journal}
  {The Journal of Physical Chemistry}\ }\textbf {\bibinfo {volume} {86}},\
  \bibinfo {pages} {3345} (\bibinfo {year} {1982})}\BibitemShut {NoStop}%
\bibitem [{\citenamefont {Thorsm{\o}lle}\ \emph {et~al.}(2009)\citenamefont
  {Thorsm{\o}lle}, \citenamefont {Averitt}, \citenamefont {Demsar},
  \citenamefont {Smith}, \citenamefont {Tretiak}, \citenamefont {Martin},
  \citenamefont {Chi}, \citenamefont {Crone}, \citenamefont {Ramirez},\ and\
  \citenamefont {Taylor}}]{art:Thorsmolle09}%
  \BibitemOpen
  \bibfield  {author} {\bibinfo {author} {\bibfnamefont {V.~K.}\ \bibnamefont
  {Thorsm{\o}lle}}, \bibinfo {author} {\bibfnamefont {R.~D.}\ \bibnamefont
  {Averitt}}, \bibinfo {author} {\bibfnamefont {J.}~\bibnamefont {Demsar}},
  \bibinfo {author} {\bibfnamefont {D.~L.}\ \bibnamefont {Smith}}, \bibinfo
  {author} {\bibfnamefont {S.}~\bibnamefont {Tretiak}}, \bibinfo {author}
  {\bibfnamefont {R.~L.}\ \bibnamefont {Martin}}, \bibinfo {author}
  {\bibfnamefont {X.}~\bibnamefont {Chi}}, \bibinfo {author} {\bibfnamefont
  {B.~K.}\ \bibnamefont {Crone}}, \bibinfo {author} {\bibfnamefont {A.~P.}\
  \bibnamefont {Ramirez}}, \ and\ \bibinfo {author} {\bibfnamefont {A.~J.}\
  \bibnamefont {Taylor}},\ }\href {\doibase 10.1103/physrevlett.102.017401}
  {\bibfield  {journal} {\bibinfo  {journal} {Physical Review Letters}\
  }\textbf {\bibinfo {volume} {102}} (\bibinfo {year} {2009}),\
  10.1103/physrevlett.102.017401}\BibitemShut {NoStop}%
\bibitem [{\citenamefont {Lottner}\ and\ \citenamefont
  {Slenczka}(2019)}]{art:Lottner19}%
  \BibitemOpen
  \bibfield  {author} {\bibinfo {author} {\bibfnamefont {E.-M.}\ \bibnamefont
  {Lottner}}\ and\ \bibinfo {author} {\bibfnamefont {A.}~\bibnamefont
  {Slenczka}},\ }\href {\doibase 10.1021/acs.jpca.9b04138} {\bibfield
  {journal} {\bibinfo  {journal} {The Journal of Physical Chemistry A}\
  }\textbf {\bibinfo {volume} {124}},\ \bibinfo {pages} {311} (\bibinfo {year}
  {2019})}\BibitemShut {NoStop}%
\bibitem [{\citenamefont {Burdett}, \citenamefont {Gosztola},\ and\
  \citenamefont {Bardeen}(2011)}]{art:Burdett11}%
  \BibitemOpen
  \bibfield  {author} {\bibinfo {author} {\bibfnamefont {J.~J.}\ \bibnamefont
  {Burdett}}, \bibinfo {author} {\bibfnamefont {D.}~\bibnamefont {Gosztola}}, \
  and\ \bibinfo {author} {\bibfnamefont {C.~J.}\ \bibnamefont {Bardeen}},\
  }\href {\doibase 10.1063/1.3664630} {\bibfield  {journal} {\bibinfo
  {journal} {The Journal of Chemical Physics}\ }\textbf {\bibinfo {volume}
  {135}},\ \bibinfo {pages} {214508} (\bibinfo {year} {2011})}\BibitemShut
  {NoStop}%
\bibitem [{\citenamefont {Müller}\ \emph {et~al.}(2015)\citenamefont
  {Müller}, \citenamefont {Izadnia}, \citenamefont {Vlaming}, \citenamefont
  {Eisfeld}, \citenamefont {LaForge},\ and\ \citenamefont
  {Stienkemeier}}]{art:Mueller15}%
  \BibitemOpen
  \bibfield  {author} {\bibinfo {author} {\bibfnamefont {M.}~\bibnamefont
  {Müller}}, \bibinfo {author} {\bibfnamefont {S.}~\bibnamefont {Izadnia}},
  \bibinfo {author} {\bibfnamefont {S.~M.}\ \bibnamefont {Vlaming}}, \bibinfo
  {author} {\bibfnamefont {A.}~\bibnamefont {Eisfeld}}, \bibinfo {author}
  {\bibfnamefont {A.}~\bibnamefont {LaForge}}, \ and\ \bibinfo {author}
  {\bibfnamefont {F.}~\bibnamefont {Stienkemeier}},\ }\href {\doibase
  10.1103/physrevb.92.121408} {\bibfield  {journal} {\bibinfo  {journal}
  {Physical Review B}\ }\textbf {\bibinfo {volume} {92}} (\bibinfo {year}
  {2015}),\ 10.1103/physrevb.92.121408}\BibitemShut {NoStop}%
\bibitem [{\citenamefont {Camposeo}\ \emph {et~al.}(2010)\citenamefont
  {Camposeo}, \citenamefont {Polo}, \citenamefont {Tavazzi}, \citenamefont
  {Silvestri}, \citenamefont {Spearman}, \citenamefont {Cingolani},\ and\
  \citenamefont {Pisignano}}]{art:Camposeo10}%
  \BibitemOpen
  \bibfield  {author} {\bibinfo {author} {\bibfnamefont {A.}~\bibnamefont
  {Camposeo}}, \bibinfo {author} {\bibfnamefont {M.}~\bibnamefont {Polo}},
  \bibinfo {author} {\bibfnamefont {S.}~\bibnamefont {Tavazzi}}, \bibinfo
  {author} {\bibfnamefont {L.}~\bibnamefont {Silvestri}}, \bibinfo {author}
  {\bibfnamefont {P.}~\bibnamefont {Spearman}}, \bibinfo {author}
  {\bibfnamefont {R.}~\bibnamefont {Cingolani}}, \ and\ \bibinfo {author}
  {\bibfnamefont {D.}~\bibnamefont {Pisignano}},\ }\href {\doibase
  10.1103/physrevb.81.033306} {\bibfield  {journal} {\bibinfo  {journal}
  {Physical Review B}\ }\textbf {\bibinfo {volume} {81}} (\bibinfo {year}
  {2010}),\ 10.1103/physrevb.81.033306}\BibitemShut {NoStop}%
\bibitem [{\citenamefont {Piland}\ and\ \citenamefont
  {Bardeen}(2015)}]{art:Piland15}%
  \BibitemOpen
  \bibfield  {author} {\bibinfo {author} {\bibfnamefont {G.~B.}\ \bibnamefont
  {Piland}}\ and\ \bibinfo {author} {\bibfnamefont {C.~J.}\ \bibnamefont
  {Bardeen}},\ }\href {\doibase 10.1021/acs.jpclett.5b00569} {\bibfield
  {journal} {\bibinfo  {journal} {The Journal of Physical Chemistry Letters}\
  }\textbf {\bibinfo {volume} {6}},\ \bibinfo {pages} {1841} (\bibinfo {year}
  {2015})}\BibitemShut {NoStop}%
\bibitem [{\citenamefont {Korovina}\ \emph {et~al.}(2016)\citenamefont
  {Korovina}, \citenamefont {Das}, \citenamefont {Nett}, \citenamefont {Feng},
  \citenamefont {Joy}, \citenamefont {Haiges}, \citenamefont {Krylov},
  \citenamefont {Bradforth},\ and\ \citenamefont {Thompson}}]{art:Korovina16}%
  \BibitemOpen
  \bibfield  {author} {\bibinfo {author} {\bibfnamefont {N.~V.}\ \bibnamefont
  {Korovina}}, \bibinfo {author} {\bibfnamefont {S.}~\bibnamefont {Das}},
  \bibinfo {author} {\bibfnamefont {Z.}~\bibnamefont {Nett}}, \bibinfo {author}
  {\bibfnamefont {X.}~\bibnamefont {Feng}}, \bibinfo {author} {\bibfnamefont
  {J.}~\bibnamefont {Joy}}, \bibinfo {author} {\bibfnamefont {R.}~\bibnamefont
  {Haiges}}, \bibinfo {author} {\bibfnamefont {A.~I.}\ \bibnamefont {Krylov}},
  \bibinfo {author} {\bibfnamefont {S.~E.}\ \bibnamefont {Bradforth}}, \ and\
  \bibinfo {author} {\bibfnamefont {M.~E.}\ \bibnamefont {Thompson}},\ }\href
  {\doibase 10.1021/jacs.5b10550} {\bibfield  {journal} {\bibinfo  {journal}
  {Journal of the American Chemical Society}\ }\textbf {\bibinfo {volume}
  {138}},\ \bibinfo {pages} {617} (\bibinfo {year} {2016})}\BibitemShut
  {NoStop}%
\bibitem [{\citenamefont {Garoni}\ \emph {et~al.}(2017)\citenamefont {Garoni},
  \citenamefont {Zirzlmeier}, \citenamefont {Basel}, \citenamefont {Hetzer},
  \citenamefont {Kamada}, \citenamefont {Guldi},\ and\ \citenamefont
  {Tykwinski}}]{art:Garoni17}%
  \BibitemOpen
  \bibfield  {author} {\bibinfo {author} {\bibfnamefont {E.}~\bibnamefont
  {Garoni}}, \bibinfo {author} {\bibfnamefont {J.}~\bibnamefont {Zirzlmeier}},
  \bibinfo {author} {\bibfnamefont {B.~S.}\ \bibnamefont {Basel}}, \bibinfo
  {author} {\bibfnamefont {C.}~\bibnamefont {Hetzer}}, \bibinfo {author}
  {\bibfnamefont {K.}~\bibnamefont {Kamada}}, \bibinfo {author} {\bibfnamefont
  {D.~M.}\ \bibnamefont {Guldi}}, \ and\ \bibinfo {author} {\bibfnamefont
  {R.~R.}\ \bibnamefont {Tykwinski}},\ }\href {\doibase 10.1021/jacs.7b08287}
  {\bibfield  {journal} {\bibinfo  {journal} {Journal of the American Chemical
  Society}\ }\textbf {\bibinfo {volume} {139}},\ \bibinfo {pages} {14017}
  (\bibinfo {year} {2017})}\BibitemShut {NoStop}%
\bibitem [{\citenamefont {Hofmann}\ \emph {et~al.}(1979)\citenamefont
  {Hofmann}, \citenamefont {Seefeld}, \citenamefont {Hofberger},\ and\
  \citenamefont {Bässler}}]{art:Hofmann79}%
  \BibitemOpen
  \bibfield  {author} {\bibinfo {author} {\bibfnamefont {J.}~\bibnamefont
  {Hofmann}}, \bibinfo {author} {\bibfnamefont {K.}~\bibnamefont {Seefeld}},
  \bibinfo {author} {\bibfnamefont {W.}~\bibnamefont {Hofberger}}, \ and\
  \bibinfo {author} {\bibfnamefont {H.}~\bibnamefont {Bässler}},\ }\href
  {\doibase 10.1080/00268977900103321} {\bibfield  {journal} {\bibinfo
  {journal} {Molecular Physics}\ }\textbf {\bibinfo {volume} {37}},\ \bibinfo
  {pages} {973} (\bibinfo {year} {1979})}\BibitemShut {NoStop}%
\bibitem [{\citenamefont {Kobayashi}, \citenamefont {Nagakura},\ and\
  \citenamefont {Szwarc}(1979)}]{art:Kobayashi79}%
  \BibitemOpen
  \bibfield  {author} {\bibinfo {author} {\bibfnamefont {T.}~\bibnamefont
  {Kobayashi}}, \bibinfo {author} {\bibfnamefont {S.}~\bibnamefont {Nagakura}},
  \ and\ \bibinfo {author} {\bibfnamefont {M.}~\bibnamefont {Szwarc}},\ }\href
  {\doibase 10.1016/0301-0104(79)85080-6} {\bibfield  {journal} {\bibinfo
  {journal} {Chemical Physics}\ }\textbf {\bibinfo {volume} {39}},\ \bibinfo
  {pages} {105} (\bibinfo {year} {1979})}\BibitemShut {NoStop}%
\bibitem [{\citenamefont {Chen}, \citenamefont {Neels},\ and\ \citenamefont
  {Fromm}(2010)}]{art:Chen10}%
  \BibitemOpen
  \bibfield  {author} {\bibinfo {author} {\bibfnamefont {J.}~\bibnamefont
  {Chen}}, \bibinfo {author} {\bibfnamefont {A.}~\bibnamefont {Neels}}, \ and\
  \bibinfo {author} {\bibfnamefont {K.~M.}\ \bibnamefont {Fromm}},\ }\href
  {\doibase 10.1039/c0cc03011b} {\bibfield  {journal} {\bibinfo  {journal}
  {Chemical Communications}\ }\textbf {\bibinfo {volume} {46}},\ \bibinfo
  {pages} {8282} (\bibinfo {year} {2010})}\BibitemShut {NoStop}%
\bibitem [{\citenamefont {Dover}\ \emph {et~al.}(2018)\citenamefont {Dover},
  \citenamefont {Gallaher}, \citenamefont {Frazer}, \citenamefont {Tapping},
  \citenamefont {Petty}, \citenamefont {Crossley}, \citenamefont {Anthony},
  \citenamefont {Kee},\ and\ \citenamefont {Schmidt}}]{art:Dover18}%
  \BibitemOpen
  \bibfield  {author} {\bibinfo {author} {\bibfnamefont {C.~B.}\ \bibnamefont
  {Dover}}, \bibinfo {author} {\bibfnamefont {J.~K.}\ \bibnamefont {Gallaher}},
  \bibinfo {author} {\bibfnamefont {L.}~\bibnamefont {Frazer}}, \bibinfo
  {author} {\bibfnamefont {P.~C.}\ \bibnamefont {Tapping}}, \bibinfo {author}
  {\bibfnamefont {A.~J.}\ \bibnamefont {Petty}}, \bibinfo {author}
  {\bibfnamefont {M.~J.}\ \bibnamefont {Crossley}}, \bibinfo {author}
  {\bibfnamefont {J.~E.}\ \bibnamefont {Anthony}}, \bibinfo {author}
  {\bibfnamefont {T.~W.}\ \bibnamefont {Kee}}, \ and\ \bibinfo {author}
  {\bibfnamefont {T.~W.}\ \bibnamefont {Schmidt}},\ }\href {\doibase
  10.1038/nchem.2926} {\bibfield  {journal} {\bibinfo  {journal} {Nature
  Chemistry}\ }\textbf {\bibinfo {volume} {10}},\ \bibinfo {pages} {305}
  (\bibinfo {year} {2018})}\BibitemShut {NoStop}%
\bibitem [{\citenamefont {Sugino}\ \emph {et~al.}(2013)\citenamefont {Sugino},
  \citenamefont {Araki}, \citenamefont {Hatanaka}, \citenamefont {Hisaki},
  \citenamefont {Miyata},\ and\ \citenamefont {Tohnai}}]{art:Sugino13}%
  \BibitemOpen
  \bibfield  {author} {\bibinfo {author} {\bibfnamefont {M.}~\bibnamefont
  {Sugino}}, \bibinfo {author} {\bibfnamefont {Y.}~\bibnamefont {Araki}},
  \bibinfo {author} {\bibfnamefont {K.}~\bibnamefont {Hatanaka}}, \bibinfo
  {author} {\bibfnamefont {I.}~\bibnamefont {Hisaki}}, \bibinfo {author}
  {\bibfnamefont {M.}~\bibnamefont {Miyata}}, \ and\ \bibinfo {author}
  {\bibfnamefont {N.}~\bibnamefont {Tohnai}},\ }\href {\doibase
  10.1021/cg401166v} {\bibfield  {journal} {\bibinfo  {journal} {Crystal Growth
  {\&} Design}\ }\textbf {\bibinfo {volume} {13}},\ \bibinfo {pages} {4986}
  (\bibinfo {year} {2013})}\BibitemShut {NoStop}%
\bibitem [{\citenamefont {Ishii}\ \emph {et~al.}(1994)\citenamefont {Ishii},
  \citenamefont {Nakayama}, \citenamefont {Yagasaki}, \citenamefont {Ando},\
  and\ \citenamefont {Kawahara}}]{art:Ishii94}%
  \BibitemOpen
  \bibfield  {author} {\bibinfo {author} {\bibfnamefont {K.}~\bibnamefont
  {Ishii}}, \bibinfo {author} {\bibfnamefont {H.}~\bibnamefont {Nakayama}},
  \bibinfo {author} {\bibfnamefont {Y.}~\bibnamefont {Yagasaki}}, \bibinfo
  {author} {\bibfnamefont {K.}~\bibnamefont {Ando}}, \ and\ \bibinfo {author}
  {\bibfnamefont {M.}~\bibnamefont {Kawahara}},\ }\href {\doibase
  10.1016/0009-2614(94)00334-3} {\bibfield  {journal} {\bibinfo  {journal}
  {Chemical Physics Letters}\ }\textbf {\bibinfo {volume} {222}},\ \bibinfo
  {pages} {117} (\bibinfo {year} {1994})}\BibitemShut {NoStop}%
\bibitem [{\citenamefont {Izadnia}\ \emph {et~al.}(2017)\citenamefont
  {Izadnia}, \citenamefont {Schönleber}, \citenamefont {Eisfeld},
  \citenamefont {Ruf}, \citenamefont {LaForge},\ and\ \citenamefont
  {Stienkemeier}}]{art:Izadnia17}%
  \BibitemOpen
  \bibfield  {author} {\bibinfo {author} {\bibfnamefont {S.}~\bibnamefont
  {Izadnia}}, \bibinfo {author} {\bibfnamefont {D.~W.}\ \bibnamefont
  {Schönleber}}, \bibinfo {author} {\bibfnamefont {A.}~\bibnamefont
  {Eisfeld}}, \bibinfo {author} {\bibfnamefont {A.}~\bibnamefont {Ruf}},
  \bibinfo {author} {\bibfnamefont {A.~C.}\ \bibnamefont {LaForge}}, \ and\
  \bibinfo {author} {\bibfnamefont {F.}~\bibnamefont {Stienkemeier}},\ }\href
  {\doibase 10.1021/acs.jpclett.7b00319} {\bibfield  {journal} {\bibinfo
  {journal} {The Journal of Physical Chemistry Letters}\ }\textbf {\bibinfo
  {volume} {8}},\ \bibinfo {pages} {2068} (\bibinfo {year} {2017})}\BibitemShut
  {NoStop}%
\bibitem [{\citenamefont {Izadnia}(2018)}]{diss:Izadnia2018}%
  \BibitemOpen
  \bibfield  {author} {\bibinfo {author} {\bibfnamefont {S.}~\bibnamefont
  {Izadnia}},\ }\emph {\bibinfo {title} {Fluorescence lifetime reduction
  mechanisms of organic complexes studied by cluster isolation spectroscopy}},\
  \href {\doibase 10.6094/UNIFR/16006} {Ph.D. thesis},\ \bibinfo  {school}
  {University of Freiburg} (\bibinfo {year} {2018})\BibitemShut {NoStop}%
\bibitem [{\citenamefont {Farges}\ \emph {et~al.}(1981)\citenamefont {Farges},
  \citenamefont {de~Feraudy}, \citenamefont {Raoult},\ and\ \citenamefont
  {Torchet}}]{art:Farges81}%
  \BibitemOpen
  \bibfield  {author} {\bibinfo {author} {\bibfnamefont {J.}~\bibnamefont
  {Farges}}, \bibinfo {author} {\bibfnamefont {M.}~\bibnamefont {de~Feraudy}},
  \bibinfo {author} {\bibfnamefont {B.}~\bibnamefont {Raoult}}, \ and\ \bibinfo
  {author} {\bibfnamefont {G.}~\bibnamefont {Torchet}},\ }\href {\doibase
  10.1016/0039-6028(81)90186-2} {\bibfield  {journal} {\bibinfo  {journal}
  {Surface Science}\ }\textbf {\bibinfo {volume} {106}},\ \bibinfo {pages} {95}
  (\bibinfo {year} {1981})}\BibitemShut {NoStop}%
\bibitem [{\citenamefont {Even}\ \emph {et~al.}(2000)\citenamefont {Even},
  \citenamefont {Jortner}, \citenamefont {Noy}, \citenamefont {Lavie},\ and\
  \citenamefont {Cossart-Magos}}]{art:Even00}%
  \BibitemOpen
  \bibfield  {author} {\bibinfo {author} {\bibfnamefont {U.}~\bibnamefont
  {Even}}, \bibinfo {author} {\bibfnamefont {J.}~\bibnamefont {Jortner}},
  \bibinfo {author} {\bibfnamefont {D.}~\bibnamefont {Noy}}, \bibinfo {author}
  {\bibfnamefont {N.}~\bibnamefont {Lavie}}, \ and\ \bibinfo {author}
  {\bibfnamefont {C.}~\bibnamefont {Cossart-Magos}},\ }\href {\doibase
  10.1063/1.481405} {\bibfield  {journal} {\bibinfo  {journal} {The Journal of
  Chemical Physics}\ }\textbf {\bibinfo {volume} {112}},\ \bibinfo {pages}
  {8068} (\bibinfo {year} {2000})}\BibitemShut {NoStop}%
\bibitem [{\citenamefont {Luria}, \citenamefont {Christen},\ and\ \citenamefont
  {Even}(2011)}]{art:Luria11}%
  \BibitemOpen
  \bibfield  {author} {\bibinfo {author} {\bibfnamefont {K.}~\bibnamefont
  {Luria}}, \bibinfo {author} {\bibfnamefont {W.}~\bibnamefont {Christen}}, \
  and\ \bibinfo {author} {\bibfnamefont {U.}~\bibnamefont {Even}},\ }\href
  {\doibase 10.1021/jp201342u} {\bibfield  {journal} {\bibinfo  {journal} {The
  Journal of Physical Chemistry A}\ }\textbf {\bibinfo {volume} {115}},\
  \bibinfo {pages} {7362} (\bibinfo {year} {2011})}\BibitemShut {NoStop}%
\bibitem [{\citenamefont {Gough}\ \emph {et~al.}(1985)\citenamefont {Gough},
  \citenamefont {Mengel}, \citenamefont {Rowntree},\ and\ \citenamefont
  {Scoles}}]{art:Gough85}%
  \BibitemOpen
  \bibfield  {author} {\bibinfo {author} {\bibfnamefont {T.~E.}\ \bibnamefont
  {Gough}}, \bibinfo {author} {\bibfnamefont {M.}~\bibnamefont {Mengel}},
  \bibinfo {author} {\bibfnamefont {P.~A.}\ \bibnamefont {Rowntree}}, \ and\
  \bibinfo {author} {\bibfnamefont {G.}~\bibnamefont {Scoles}},\ }\href
  {\doibase 10.1063/1.449757} {\bibfield  {journal} {\bibinfo  {journal} {The
  Journal of Chemical Physics}\ }\textbf {\bibinfo {volume} {83}},\ \bibinfo
  {pages} {4958} (\bibinfo {year} {1985})}\BibitemShut {NoStop}%
\bibitem [{\citenamefont {Hagena}(1992)}]{art:Hagena92}%
  \BibitemOpen
  \bibfield  {author} {\bibinfo {author} {\bibfnamefont {O.~F.}\ \bibnamefont
  {Hagena}},\ }\href {\doibase 10.1063/1.1142933} {\bibfield  {journal}
  {\bibinfo  {journal} {Review of Scientific Instruments}\ }\textbf {\bibinfo
  {volume} {63}},\ \bibinfo {pages} {2374} (\bibinfo {year}
  {1992})}\BibitemShut {NoStop}%
\bibitem [{\citenamefont {Buck}\ and\ \citenamefont
  {Krohne}(1996)}]{art:Buck96}%
  \BibitemOpen
  \bibfield  {author} {\bibinfo {author} {\bibfnamefont {U.}~\bibnamefont
  {Buck}}\ and\ \bibinfo {author} {\bibfnamefont {R.}~\bibnamefont {Krohne}},\
  }\href {\doibase http://dx.doi.org/10.1063/1.472406} {\bibfield  {journal}
  {\bibinfo  {journal} {The Journal of Chemical Physics}\ }\textbf {\bibinfo
  {volume} {105}},\ \bibinfo {pages} {5408} (\bibinfo {year}
  {1996})}\BibitemShut {NoStop}%
\bibitem [{\citenamefont {Dorchies}\ \emph {et~al.}(2003)\citenamefont
  {Dorchies}, \citenamefont {Blasco}, \citenamefont {Caillaud}, \citenamefont
  {Stevefelt}, \citenamefont {Stenz}, \citenamefont {Boldarev},\ and\
  \citenamefont {Gasilov}}]{art:Dorchies03}%
  \BibitemOpen
  \bibfield  {author} {\bibinfo {author} {\bibfnamefont {F.}~\bibnamefont
  {Dorchies}}, \bibinfo {author} {\bibfnamefont {F.}~\bibnamefont {Blasco}},
  \bibinfo {author} {\bibfnamefont {T.}~\bibnamefont {Caillaud}}, \bibinfo
  {author} {\bibfnamefont {J.}~\bibnamefont {Stevefelt}}, \bibinfo {author}
  {\bibfnamefont {C.}~\bibnamefont {Stenz}}, \bibinfo {author} {\bibfnamefont
  {A.~S.}\ \bibnamefont {Boldarev}}, \ and\ \bibinfo {author} {\bibfnamefont
  {V.~A.}\ \bibnamefont {Gasilov}},\ }\href {\doibase
  10.1103/physreva.68.023201} {\bibfield  {journal} {\bibinfo  {journal}
  {Physical Review A}\ }\textbf {\bibinfo {volume} {68}} (\bibinfo {year}
  {2003}),\ 10.1103/physreva.68.023201}\BibitemShut {NoStop}%
\bibitem [{\citenamefont {Lee}\ and\ \citenamefont {Stein}(1987)}]{art:Lee87}%
  \BibitemOpen
  \bibfield  {author} {\bibinfo {author} {\bibfnamefont {J.~W.}\ \bibnamefont
  {Lee}}\ and\ \bibinfo {author} {\bibfnamefont {G.~D.}\ \bibnamefont
  {Stein}},\ }\href {\doibase 10.1021/j100294a001} {\bibfield  {journal}
  {\bibinfo  {journal} {The Journal of Physical Chemistry}\ }\textbf {\bibinfo
  {volume} {91}},\ \bibinfo {pages} {2450} (\bibinfo {year}
  {1987})}\BibitemShut {NoStop}%
\bibitem [{\citenamefont {Pauly}(2000)}]{book:Pauly2000}%
  \BibitemOpen
  \bibfield  {author} {\bibinfo {author} {\bibfnamefont {H.}~\bibnamefont
  {Pauly}},\ }\href
  {https://www.ebook.de/de/product/6699434/hans_pauly_atom_molecule_and_cluster_beams_i.html}
  {\emph {\bibinfo {title} {Atom, Molecule, and Cluster Beams I}}}\ (\bibinfo
  {publisher} {Springer Berlin Heidelberg},\ \bibinfo {year}
  {2000})\BibitemShut {NoStop}%
\bibitem [{\citenamefont {Ferreira}\ and\ \citenamefont
  {Lobo}(2008)}]{art:Ferreira08}%
  \BibitemOpen
  \bibfield  {author} {\bibinfo {author} {\bibfnamefont {A.}~\bibnamefont
  {Ferreira}}\ and\ \bibinfo {author} {\bibfnamefont {L.}~\bibnamefont
  {Lobo}},\ }\href {\doibase 10.1016/j.jct.2008.07.023} {\bibfield  {journal}
  {\bibinfo  {journal} {The Journal of Chemical Thermodynamics}\ }\textbf
  {\bibinfo {volume} {40}},\ \bibinfo {pages} {1621} (\bibinfo {year}
  {2008})}\BibitemShut {NoStop}%
\bibitem [{\citenamefont {Suuberg}\ and\ \citenamefont
  {Oja}(1997)}]{techrep:Suuberg97}%
  \BibitemOpen
  \bibfield  {author} {\bibinfo {author} {\bibfnamefont {E.~M.}\ \bibnamefont
  {Suuberg}}\ and\ \bibinfo {author} {\bibfnamefont {V.}~\bibnamefont {Oja}},\
  }\href {\doibase 10.2172/774960} {\enquote {\bibinfo {title} {{VAPOR}
  {PRESSURES} {AND} {HEATS} {OF} {VAPORIZATION} {OF} {PRIMARY} {COAL}
  {TARS}},}\ }\bibinfo {type} {Tech. Rep.}\ (\bibinfo  {institution} {Federal
  Energy Technology Center Morgantown (FETC-MGN), Morgantown, WV (United
  States); Federal Energy Technology Center Pittsburgh (FETC-PGH), Pittsburgh,
  PA (United States)},\ \bibinfo {year} {1997})\BibitemShut {NoStop}%
\bibitem [{\citenamefont {Stephenson}\ and\ \citenamefont
  {Malanowski}(1987)}]{book:Stephenson1987}%
  \BibitemOpen
  \bibfield  {author} {\bibinfo {author} {\bibfnamefont {R.~M.}\ \bibnamefont
  {Stephenson}}\ and\ \bibinfo {author} {\bibfnamefont {S.}~\bibnamefont
  {Malanowski}},\ }\href {\doibase 10.1007/978-94-009-3173-2_1} {\emph
  {\bibinfo {title} {Properties of Organic Compounds}}}\ (\bibinfo  {publisher}
  {Springer Netherlands},\ \bibinfo {year} {1987})\ pp.\ \bibinfo {pages}
  {1--471}\BibitemShut {NoStop}%
\bibitem [{\citenamefont {Scott}\ and\ \citenamefont
  {Tout}(1989)}]{art:Scott89}%
  \BibitemOpen
  \bibfield  {author} {\bibinfo {author} {\bibfnamefont {D.}~\bibnamefont
  {Scott}}\ and\ \bibinfo {author} {\bibfnamefont {C.~A.}\ \bibnamefont
  {Tout}},\ }\href {\doibase 10.1093/mnras/241.2.109} {\bibfield  {journal}
  {\bibinfo  {journal} {Monthly Notices of the Royal Astronomical Society}\
  }\textbf {\bibinfo {volume} {241}},\ \bibinfo {pages} {109} (\bibinfo {year}
  {1989})}\BibitemShut {NoStop}%
\bibitem [{\citenamefont {Mortensen}, \citenamefont {Hansen},\ and\
  \citenamefont {Jacobsen}(2005)}]{art:Mortensen05}%
  \BibitemOpen
  \bibfield  {author} {\bibinfo {author} {\bibfnamefont {J.~J.}\ \bibnamefont
  {Mortensen}}, \bibinfo {author} {\bibfnamefont {L.~B.}\ \bibnamefont
  {Hansen}}, \ and\ \bibinfo {author} {\bibfnamefont {K.~W.}\ \bibnamefont
  {Jacobsen}},\ }\href {\doibase 10.1103/PhysRevB.71.035109} {\bibfield
  {journal} {\bibinfo  {journal} {Physical Review B}\ }\textbf {\bibinfo
  {volume} {71}},\ \bibinfo {pages} {035109} (\bibinfo {year}
  {2005})}\BibitemShut {NoStop}%
\bibitem [{\citenamefont {Enkovaara}\ \emph {et~al.}(2010)\citenamefont
  {Enkovaara}, \citenamefont {Rostgaard}, \citenamefont {Mortensen},
  \citenamefont {Chen}, \citenamefont {Dułak}, \citenamefont {Ferrighi},
  \citenamefont {Gavnholt}, \citenamefont {Glinsvad}, \citenamefont {Haikola},
  \citenamefont {Hansen}, \citenamefont {Kristoffersen}, \citenamefont
  {Kuisma}, \citenamefont {Larsen}, \citenamefont {Lehtovaara}, \citenamefont
  {Ljungberg}, \citenamefont {Lopez-Acevedo}, \citenamefont {Moses},
  \citenamefont {Ojanen}, \citenamefont {Olsen}, \citenamefont {Petzold},
  \citenamefont {Romero}, \citenamefont {Stausholm-Møller}, \citenamefont
  {Strange}, \citenamefont {Tritsaris}, \citenamefont {Vanin}, \citenamefont
  {Walter}, \citenamefont {Hammer}, \citenamefont {Häkkinen}, \citenamefont
  {Madsen}, \citenamefont {Nieminen}, \citenamefont {Nørskov}, \citenamefont
  {Puska}, \citenamefont {Rantala}, \citenamefont {Schiøtz}, \citenamefont
  {Thygesen},\ and\ \citenamefont {Jacobsen}}]{art:Enkovaara10}%
  \BibitemOpen
  \bibfield  {author} {\bibinfo {author} {\bibfnamefont {J.}~\bibnamefont
  {Enkovaara}}, \bibinfo {author} {\bibfnamefont {C.}~\bibnamefont
  {Rostgaard}}, \bibinfo {author} {\bibfnamefont {J.~J.}\ \bibnamefont
  {Mortensen}}, \bibinfo {author} {\bibfnamefont {J.}~\bibnamefont {Chen}},
  \bibinfo {author} {\bibfnamefont {M.}~\bibnamefont {Dułak}}, \bibinfo
  {author} {\bibfnamefont {L.}~\bibnamefont {Ferrighi}}, \bibinfo {author}
  {\bibfnamefont {J.}~\bibnamefont {Gavnholt}}, \bibinfo {author}
  {\bibfnamefont {C.}~\bibnamefont {Glinsvad}}, \bibinfo {author}
  {\bibfnamefont {V.}~\bibnamefont {Haikola}}, \bibinfo {author} {\bibfnamefont
  {H.~A.}\ \bibnamefont {Hansen}}, \bibinfo {author} {\bibfnamefont {H.~H.}\
  \bibnamefont {Kristoffersen}}, \bibinfo {author} {\bibfnamefont
  {M.}~\bibnamefont {Kuisma}}, \bibinfo {author} {\bibfnamefont {A.~H.}\
  \bibnamefont {Larsen}}, \bibinfo {author} {\bibfnamefont {L.}~\bibnamefont
  {Lehtovaara}}, \bibinfo {author} {\bibfnamefont {M.}~\bibnamefont
  {Ljungberg}}, \bibinfo {author} {\bibfnamefont {O.}~\bibnamefont
  {Lopez-Acevedo}}, \bibinfo {author} {\bibfnamefont {P.~G.}\ \bibnamefont
  {Moses}}, \bibinfo {author} {\bibfnamefont {J.}~\bibnamefont {Ojanen}},
  \bibinfo {author} {\bibfnamefont {T.}~\bibnamefont {Olsen}}, \bibinfo
  {author} {\bibfnamefont {V.}~\bibnamefont {Petzold}}, \bibinfo {author}
  {\bibfnamefont {N.~A.}\ \bibnamefont {Romero}}, \bibinfo {author}
  {\bibfnamefont {J.}~\bibnamefont {Stausholm-Møller}}, \bibinfo {author}
  {\bibfnamefont {M.}~\bibnamefont {Strange}}, \bibinfo {author} {\bibfnamefont
  {G.~A.}\ \bibnamefont {Tritsaris}}, \bibinfo {author} {\bibfnamefont
  {M.}~\bibnamefont {Vanin}}, \bibinfo {author} {\bibfnamefont
  {M.}~\bibnamefont {Walter}}, \bibinfo {author} {\bibfnamefont
  {B.}~\bibnamefont {Hammer}}, \bibinfo {author} {\bibfnamefont
  {H.}~\bibnamefont {Häkkinen}}, \bibinfo {author} {\bibfnamefont {G.~K.~H.}\
  \bibnamefont {Madsen}}, \bibinfo {author} {\bibfnamefont {R.~M.}\
  \bibnamefont {Nieminen}}, \bibinfo {author} {\bibfnamefont {J.~K.}\
  \bibnamefont {Nørskov}}, \bibinfo {author} {\bibfnamefont {M.}~\bibnamefont
  {Puska}}, \bibinfo {author} {\bibfnamefont {T.~T.}\ \bibnamefont {Rantala}},
  \bibinfo {author} {\bibfnamefont {J.}~\bibnamefont {Schiøtz}}, \bibinfo
  {author} {\bibfnamefont {K.~S.}\ \bibnamefont {Thygesen}}, \ and\ \bibinfo
  {author} {\bibfnamefont {K.~W.}\ \bibnamefont {Jacobsen}},\ }\href {\doibase
  10.1088/0953-8984/22/25/253202} {\bibfield  {journal} {\bibinfo  {journal}
  {Journal of Physics: Condensed Matter}\ }\textbf {\bibinfo {volume} {22}},\
  \bibinfo {pages} {253202} (\bibinfo {year} {2010})}\BibitemShut {NoStop}%
\bibitem [{\citenamefont {Walter}\ \emph {et~al.}(2008)\citenamefont {Walter},
  \citenamefont {Häkkinen}, \citenamefont {Lehtovaara}, \citenamefont {Puska},
  \citenamefont {Enkovaara}, \citenamefont {Rostgaard},\ and\ \citenamefont
  {Mortensen}}]{art:Walter08}%
  \BibitemOpen
  \bibfield  {author} {\bibinfo {author} {\bibfnamefont {M.}~\bibnamefont
  {Walter}}, \bibinfo {author} {\bibfnamefont {H.}~\bibnamefont {Häkkinen}},
  \bibinfo {author} {\bibfnamefont {L.}~\bibnamefont {Lehtovaara}}, \bibinfo
  {author} {\bibfnamefont {M.}~\bibnamefont {Puska}}, \bibinfo {author}
  {\bibfnamefont {J.}~\bibnamefont {Enkovaara}}, \bibinfo {author}
  {\bibfnamefont {C.}~\bibnamefont {Rostgaard}}, \ and\ \bibinfo {author}
  {\bibfnamefont {J.~J.}\ \bibnamefont {Mortensen}},\ }\href {\doibase
  10.1063/1.2943138} {\bibfield  {journal} {\bibinfo  {journal} {The Journal of
  Chemical Physics}\ }\textbf {\bibinfo {volume} {128}},\ \bibinfo {pages}
  {244101} (\bibinfo {year} {2008})}\BibitemShut {NoStop}%
\bibitem [{\citenamefont {Stauffert}\ \emph {et~al.}(2019)\citenamefont
  {Stauffert}, \citenamefont {Izadnia}, \citenamefont {Stienkemeier},\ and\
  \citenamefont {Walter}}]{art:Stauffert19}%
  \BibitemOpen
  \bibfield  {author} {\bibinfo {author} {\bibfnamefont {O.}~\bibnamefont
  {Stauffert}}, \bibinfo {author} {\bibfnamefont {S.}~\bibnamefont {Izadnia}},
  \bibinfo {author} {\bibfnamefont {F.}~\bibnamefont {Stienkemeier}}, \ and\
  \bibinfo {author} {\bibfnamefont {M.}~\bibnamefont {Walter}},\ }\href
  {\doibase 10.1063/1.5097553} {\bibfield  {journal} {\bibinfo  {journal} {The
  Journal of Chemical Physics}\ }\textbf {\bibinfo {volume} {150}},\ \bibinfo
  {pages} {244703} (\bibinfo {year} {2019})}\BibitemShut {NoStop}%
\bibitem [{\citenamefont {Lambert}\ \emph {et~al.}(1984)\citenamefont
  {Lambert}, \citenamefont {Felker}, \citenamefont {Syage},\ and\ \citenamefont
  {Zewail}}]{art:Lambert84}%
  \BibitemOpen
  \bibfield  {author} {\bibinfo {author} {\bibfnamefont {W.~R.}\ \bibnamefont
  {Lambert}}, \bibinfo {author} {\bibfnamefont {P.~M.}\ \bibnamefont {Felker}},
  \bibinfo {author} {\bibfnamefont {J.~A.}\ \bibnamefont {Syage}}, \ and\
  \bibinfo {author} {\bibfnamefont {A.~H.}\ \bibnamefont {Zewail}},\ }\href
  {\doibase 10.1063/1.447922} {\bibfield  {journal} {\bibinfo  {journal} {The
  Journal of Chemical Physics}\ }\textbf {\bibinfo {volume} {81}},\ \bibinfo
  {pages} {2195} (\bibinfo {year} {1984})}\BibitemShut {NoStop}%
\bibitem [{\citenamefont {Zalesskaya}\ \emph {et~al.}(2002)\citenamefont
  {Zalesskaya}, \citenamefont {Pavlova}, \citenamefont {Yakovlev},
  \citenamefont {Sambor},\ and\ \citenamefont {Belyi}}]{art:Zalesskaya02}%
  \BibitemOpen
  \bibfield  {author} {\bibinfo {author} {\bibfnamefont {G.~A.}\ \bibnamefont
  {Zalesskaya}}, \bibinfo {author} {\bibfnamefont {V.~T.}\ \bibnamefont
  {Pavlova}}, \bibinfo {author} {\bibfnamefont {D.~L.}\ \bibnamefont
  {Yakovlev}}, \bibinfo {author} {\bibfnamefont {E.~G.}\ \bibnamefont
  {Sambor}}, \ and\ \bibinfo {author} {\bibfnamefont {N.~N.}\ \bibnamefont
  {Belyi}},\ }\href {\doibase 10.1134/1.1531707} {\bibfield  {journal}
  {\bibinfo  {journal} {Optics and Spectroscopy}\ }\textbf {\bibinfo {volume}
  {93}},\ \bibinfo {pages} {848} (\bibinfo {year} {2002})}\BibitemShut
  {NoStop}%
\bibitem [{\citenamefont {Pentlehner}\ \emph {et~al.}(2010)\citenamefont
  {Pentlehner}, \citenamefont {Greil}, \citenamefont {Dick},\ and\
  \citenamefont {Slenczka}}]{art:Pentlehner10}%
  \BibitemOpen
  \bibfield  {author} {\bibinfo {author} {\bibfnamefont {D.}~\bibnamefont
  {Pentlehner}}, \bibinfo {author} {\bibfnamefont {C.}~\bibnamefont {Greil}},
  \bibinfo {author} {\bibfnamefont {B.}~\bibnamefont {Dick}}, \ and\ \bibinfo
  {author} {\bibfnamefont {A.}~\bibnamefont {Slenczka}},\ }\href {\doibase
  10.1063/1.3479583} {\bibfield  {journal} {\bibinfo  {journal} {The Journal of
  Chemical Physics}\ }\textbf {\bibinfo {volume} {133}},\ \bibinfo {pages}
  {114505} (\bibinfo {year} {2010})}\BibitemShut {NoStop}%
\bibitem [{\citenamefont {Hartmann}\ \emph {et~al.}(2001)\citenamefont
  {Hartmann}, \citenamefont {Lindinger}, \citenamefont {Toennies},\ and\
  \citenamefont {Vilesov}}]{art:Hartmann01}%
  \BibitemOpen
  \bibfield  {author} {\bibinfo {author} {\bibfnamefont {M.}~\bibnamefont
  {Hartmann}}, \bibinfo {author} {\bibfnamefont {A.}~\bibnamefont {Lindinger}},
  \bibinfo {author} {\bibfnamefont {J.~P.}\ \bibnamefont {Toennies}}, \ and\
  \bibinfo {author} {\bibfnamefont {A.~F.}\ \bibnamefont {Vilesov}},\ }\href
  {\doibase 10.1021/jp003600t} {\bibfield  {journal} {\bibinfo  {journal} {The
  Journal of Physical Chemistry A}\ }\textbf {\bibinfo {volume} {105}},\
  \bibinfo {pages} {6369} (\bibinfo {year} {2001})}\BibitemShut {NoStop}%
\bibitem [{\citenamefont {Yang}, \citenamefont {Davidson},\ and\ \citenamefont
  {Yang}(2016)}]{art:Yang16}%
  \BibitemOpen
  \bibfield  {author} {\bibinfo {author} {\bibfnamefont {Y.}~\bibnamefont
  {Yang}}, \bibinfo {author} {\bibfnamefont {E.~R.}\ \bibnamefont {Davidson}},
  \ and\ \bibinfo {author} {\bibfnamefont {W.}~\bibnamefont {Yang}},\ }\href
  {\doibase 10.1073/pnas.1606021113} {\bibfield  {journal} {\bibinfo  {journal}
  {Proceedings of the National Academy of Sciences}\ }\textbf {\bibinfo
  {volume} {113}},\ \bibinfo {pages} {E5098} (\bibinfo {year}
  {2016})}\BibitemShut {NoStop}%
\bibitem [{\citenamefont {Amirav}, \citenamefont {Even},\ and\ \citenamefont
  {Jortner}(1980)}]{art:Amirav80}%
  \BibitemOpen
  \bibfield  {author} {\bibinfo {author} {\bibfnamefont {A.}~\bibnamefont
  {Amirav}}, \bibinfo {author} {\bibfnamefont {U.}~\bibnamefont {Even}}, \ and\
  \bibinfo {author} {\bibfnamefont {J.}~\bibnamefont {Jortner}},\ }\href
  {\doibase 10.1016/0009-2614(80)80232-6} {\bibfield  {journal} {\bibinfo
  {journal} {Chemical Physics Letters}\ }\textbf {\bibinfo {volume} {72}},\
  \bibinfo {pages} {21} (\bibinfo {year} {1980})}\BibitemShut {NoStop}%
\bibitem [{\citenamefont {Lindinger}, \citenamefont {Toennies},\ and\
  \citenamefont {Vilesov}(2006)}]{art:Lindinger06}%
  \BibitemOpen
  \bibfield  {author} {\bibinfo {author} {\bibfnamefont {A.}~\bibnamefont
  {Lindinger}}, \bibinfo {author} {\bibfnamefont {J.~P.}\ \bibnamefont
  {Toennies}}, \ and\ \bibinfo {author} {\bibfnamefont {A.~F.}\ \bibnamefont
  {Vilesov}},\ }\href {\doibase 10.1016/j.cplett.2006.07.072} {\bibfield
  {journal} {\bibinfo  {journal} {Chemical Physics Letters}\ }\textbf {\bibinfo
  {volume} {429}},\ \bibinfo {pages} {1} (\bibinfo {year} {2006})}\BibitemShut
  {NoStop}%
\bibitem [{\citenamefont {Dick}, \citenamefont {Zinghar},\ and\ \citenamefont
  {Haas}(1991)}]{art:Dick91}%
  \BibitemOpen
  \bibfield  {author} {\bibinfo {author} {\bibfnamefont {B.}~\bibnamefont
  {Dick}}, \bibinfo {author} {\bibfnamefont {E.}~\bibnamefont {Zinghar}}, \
  and\ \bibinfo {author} {\bibfnamefont {Y.}~\bibnamefont {Haas}},\ }\href
  {\doibase 10.1016/0009-2614(91)90438-f} {\bibfield  {journal} {\bibinfo
  {journal} {Chemical Physics Letters}\ }\textbf {\bibinfo {volume} {187}},\
  \bibinfo {pages} {571} (\bibinfo {year} {1991})}\BibitemShut {NoStop}%
\bibitem [{\citenamefont {Halasinski}\ \emph {et~al.}(2000)\citenamefont
  {Halasinski}, \citenamefont {Hudgins}, \citenamefont {Salama}, \citenamefont
  {Allamandola},\ and\ \citenamefont {Bally}}]{art:Halasinski00}%
  \BibitemOpen
  \bibfield  {author} {\bibinfo {author} {\bibfnamefont {T.~M.}\ \bibnamefont
  {Halasinski}}, \bibinfo {author} {\bibfnamefont {D.~M.}\ \bibnamefont
  {Hudgins}}, \bibinfo {author} {\bibfnamefont {F.}~\bibnamefont {Salama}},
  \bibinfo {author} {\bibfnamefont {L.~J.}\ \bibnamefont {Allamandola}}, \ and\
  \bibinfo {author} {\bibfnamefont {T.}~\bibnamefont {Bally}},\ }\href
  {\doibase 10.1021/jp0011544} {\bibfield  {journal} {\bibinfo  {journal} {The
  Journal of Physical Chemistry A}\ }\textbf {\bibinfo {volume} {104}},\
  \bibinfo {pages} {7484} (\bibinfo {year} {2000})}\BibitemShut {NoStop}%
\bibitem [{\citenamefont {Heinecke}\ \emph {et~al.}(1998)\citenamefont
  {Heinecke}, \citenamefont {Hartmann}, \citenamefont {Müller},\ and\
  \citenamefont {Hese}}]{art:Heinecke98}%
  \BibitemOpen
  \bibfield  {author} {\bibinfo {author} {\bibfnamefont {E.}~\bibnamefont
  {Heinecke}}, \bibinfo {author} {\bibfnamefont {D.}~\bibnamefont {Hartmann}},
  \bibinfo {author} {\bibfnamefont {R.}~\bibnamefont {Müller}}, \ and\
  \bibinfo {author} {\bibfnamefont {A.}~\bibnamefont {Hese}},\ }\href {\doibase
  10.1063/1.476631} {\bibfield  {journal} {\bibinfo  {journal} {The Journal of
  Chemical Physics}\ }\textbf {\bibinfo {volume} {109}},\ \bibinfo {pages}
  {906} (\bibinfo {year} {1998})}\BibitemShut {NoStop}%
\bibitem [{\citenamefont {Nijegorodov}, \citenamefont {Ramachandran},\ and\
  \citenamefont {Winkoun}(1997)}]{art:Nijegorodov97}%
  \BibitemOpen
  \bibfield  {author} {\bibinfo {author} {\bibfnamefont {N.}~\bibnamefont
  {Nijegorodov}}, \bibinfo {author} {\bibfnamefont {V.}~\bibnamefont
  {Ramachandran}}, \ and\ \bibinfo {author} {\bibfnamefont {D.}~\bibnamefont
  {Winkoun}},\ }\href {\doibase 10.1016/s1386-1425(97)00071-1} {\bibfield
  {journal} {\bibinfo  {journal} {Spectrochimica Acta Part A: Molecular and
  Biomolecular Spectroscopy}\ }\textbf {\bibinfo {volume} {53}},\ \bibinfo
  {pages} {1813} (\bibinfo {year} {1997})}\BibitemShut {NoStop}%
\bibitem [{\citenamefont {Eisfeld}\ \emph {et~al.}(2017)\citenamefont
  {Eisfeld}, \citenamefont {Marquardt}, \citenamefont {Paulheim},\ and\
  \citenamefont {Sokolowski}}]{art:Eisfeld17}%
  \BibitemOpen
  \bibfield  {author} {\bibinfo {author} {\bibfnamefont {A.}~\bibnamefont
  {Eisfeld}}, \bibinfo {author} {\bibfnamefont {C.}~\bibnamefont {Marquardt}},
  \bibinfo {author} {\bibfnamefont {A.}~\bibnamefont {Paulheim}}, \ and\
  \bibinfo {author} {\bibfnamefont {M.}~\bibnamefont {Sokolowski}},\ }\href
  {\doibase 10.1103/physrevlett.119.097402} {\bibfield  {journal} {\bibinfo
  {journal} {Physical Review Letters}\ }\textbf {\bibinfo {volume} {119}}
  (\bibinfo {year} {2017}),\ 10.1103/physrevlett.119.097402}\BibitemShut
  {NoStop}%
\bibitem [{\citenamefont {May}\ and\ \citenamefont
  {K{\"u}hn}(2011)}]{book:May2011}%
  \BibitemOpen
  \bibfield  {author} {\bibinfo {author} {\bibfnamefont {V.}~\bibnamefont
  {May}}\ and\ \bibinfo {author} {\bibfnamefont {O.}~\bibnamefont {K{\"u}hn}},\
  }\href@noop {} {\emph {\bibinfo {title} {{Charge and Energy Transfer Dynamics
  in Molecular Systems}}}},\ \bibinfo {edition} {3rd}\ ed.\ (\bibinfo
  {publisher} {WILEY-VCH},\ \bibinfo {year} {2011})\BibitemShut {NoStop}%
\bibitem [{\citenamefont {Amirav}\ and\ \citenamefont
  {Jortner}(1986)}]{art:Amirav86}%
  \BibitemOpen
  \bibfield  {author} {\bibinfo {author} {\bibfnamefont {A.}~\bibnamefont
  {Amirav}}\ and\ \bibinfo {author} {\bibfnamefont {J.}~\bibnamefont
  {Jortner}},\ }\href {\doibase 10.1016/0009-2614(86)80621-2} {\bibfield
  {journal} {\bibinfo  {journal} {Chemical Physics Letters}\ }\textbf {\bibinfo
  {volume} {132}},\ \bibinfo {pages} {335} (\bibinfo {year}
  {1986})}\BibitemShut {NoStop}%
\bibitem [{\citenamefont {Celestino}\ and\ \citenamefont
  {Eisfeld}(2017)}]{art:Celestino17}%
  \BibitemOpen
  \bibfield  {author} {\bibinfo {author} {\bibfnamefont {A.}~\bibnamefont
  {Celestino}}\ and\ \bibinfo {author} {\bibfnamefont {A.}~\bibnamefont
  {Eisfeld}},\ }\href {\doibase 10.1021/acs.jpca.7b06259} {\bibfield  {journal}
  {\bibinfo  {journal} {The Journal of Physical Chemistry A}\ }\textbf
  {\bibinfo {volume} {121}},\ \bibinfo {pages} {5948} (\bibinfo {year}
  {2017})}\BibitemShut {NoStop}%
\bibitem [{\citenamefont {Gross}\ and\ \citenamefont
  {Haroche}(1982)}]{art:Gross82}%
  \BibitemOpen
  \bibfield  {author} {\bibinfo {author} {\bibfnamefont {M.}~\bibnamefont
  {Gross}}\ and\ \bibinfo {author} {\bibfnamefont {S.}~\bibnamefont
  {Haroche}},\ }\href {\doibase 10.1016/0370-1573(82)90102-8} {\bibfield
  {journal} {\bibinfo  {journal} {Physics Reports}\ }\textbf {\bibinfo {volume}
  {93}},\ \bibinfo {pages} {301} (\bibinfo {year} {1982})}\BibitemShut
  {NoStop}%
\bibitem [{\citenamefont {Hilborn}(1982)}]{art:Hilborn82}%
  \BibitemOpen
  \bibfield  {author} {\bibinfo {author} {\bibfnamefont {R.~C.}\ \bibnamefont
  {Hilborn}},\ }\href {\doibase 10.1119/1.12937} {\bibfield  {journal}
  {\bibinfo  {journal} {American Journal of Physics}\ }\textbf {\bibinfo
  {volume} {50}},\ \bibinfo {pages} {982} (\bibinfo {year} {1982})},\ \bibinfo
  {note} {eq. (33), extended by the refractive index to account for the reduced
  speed of light in a medium}\BibitemShut {NoStop}%
\bibitem [{\citenamefont {Schulze}\ and\ \citenamefont
  {Kolb}(1974)}]{art:Schulze74}%
  \BibitemOpen
  \bibfield  {author} {\bibinfo {author} {\bibfnamefont {W.}~\bibnamefont
  {Schulze}}\ and\ \bibinfo {author} {\bibfnamefont {D.~M.}\ \bibnamefont
  {Kolb}},\ }\href {\doibase 10.1039/f29747001098} {\bibfield  {journal}
  {\bibinfo  {journal} {Journal of the Chemical Society, Faraday Transactions
  2}\ }\textbf {\bibinfo {volume} {70}},\ \bibinfo {pages} {1098} (\bibinfo
  {year} {1974})}\BibitemShut {NoStop}%
\bibitem [{\citenamefont {Chew}(1987)}]{art:Chew87}%
  \BibitemOpen
  \bibfield  {author} {\bibinfo {author} {\bibfnamefont {H.}~\bibnamefont
  {Chew}},\ }\href {\doibase 10.1063/1.453317} {\bibfield  {journal} {\bibinfo
  {journal} {The Journal of Chemical Physics}\ }\textbf {\bibinfo {volume}
  {87}},\ \bibinfo {pages} {1355} (\bibinfo {year} {1987})}\BibitemShut
  {NoStop}%
\bibitem [{Note2()}]{Note2}%
  \BibitemOpen
  \bibinfo {note} {In Ref.~\protect \citenum {art:Nijegorodov97}, it was found
  that the rate constant for IC of Ac is more than one order of magnitude
  smaller compared to Tc and more than two orders of magnitude smaller compared
  to Pc, where it is the dominant decay channel}\BibitemShut {NoStop}%
\bibitem [{\citenamefont {Widman}\ and\ \citenamefont
  {Huber}(1972)}]{art:Widman72}%
  \BibitemOpen
  \bibfield  {author} {\bibinfo {author} {\bibfnamefont {R.~P.}\ \bibnamefont
  {Widman}}\ and\ \bibinfo {author} {\bibfnamefont {J.~R.}\ \bibnamefont
  {Huber}},\ }\href {\doibase 10.1021/j100655a005} {\bibfield  {journal}
  {\bibinfo  {journal} {The Journal of Physical Chemistry}\ }\textbf {\bibinfo
  {volume} {76}},\ \bibinfo {pages} {1524} (\bibinfo {year}
  {1972})}\BibitemShut {NoStop}%
\bibitem [{\citenamefont {Roden}\ \emph {et~al.}(2011)\citenamefont {Roden},
  \citenamefont {Eisfeld}, \citenamefont {Dvo{\v{r}}{\'{a}}k}, \citenamefont
  {Bünermann},\ and\ \citenamefont {Stienkemeier}}]{art:Roden11}%
  \BibitemOpen
  \bibfield  {author} {\bibinfo {author} {\bibfnamefont {J.}~\bibnamefont
  {Roden}}, \bibinfo {author} {\bibfnamefont {A.}~\bibnamefont {Eisfeld}},
  \bibinfo {author} {\bibfnamefont {M.}~\bibnamefont {Dvo{\v{r}}{\'{a}}k}},
  \bibinfo {author} {\bibfnamefont {O.}~\bibnamefont {Bünermann}}, \ and\
  \bibinfo {author} {\bibfnamefont {F.}~\bibnamefont {Stienkemeier}},\ }\href
  {\doibase 10.1063/1.3526749} {\bibfield  {journal} {\bibinfo  {journal} {The
  Journal of Chemical Physics}\ }\textbf {\bibinfo {volume} {134}},\ \bibinfo
  {pages} {054907} (\bibinfo {year} {2011})}\BibitemShut {NoStop}%
\bibitem [{\citenamefont {Abumwis}, \citenamefont {Eiles},\ and\ \citenamefont
  {Eisfeld}(2020{\natexlab{a}})}]{art:Abumwis20}%
  \BibitemOpen
  \bibfield  {author} {\bibinfo {author} {\bibfnamefont {G.}~\bibnamefont
  {Abumwis}}, \bibinfo {author} {\bibfnamefont {M.~T.}\ \bibnamefont {Eiles}},
  \ and\ \bibinfo {author} {\bibfnamefont {A.}~\bibnamefont {Eisfeld}},\ }\href
  {\doibase 10.1088/1361-6455/ab78a9} {\bibfield  {journal} {\bibinfo
  {journal} {Journal of Physics B: Atomic, Molecular and Optical Physics}\
  }\textbf {\bibinfo {volume} {53}},\ \bibinfo {pages} {124003} (\bibinfo
  {year} {2020}{\natexlab{a}})}\BibitemShut {NoStop}%
\bibitem [{\citenamefont {Abumwis}, \citenamefont {Eiles},\ and\ \citenamefont
  {Eisfeld}(2020{\natexlab{b}})}]{art:Abumwis20a}%
  \BibitemOpen
  \bibfield  {author} {\bibinfo {author} {\bibfnamefont {G.}~\bibnamefont
  {Abumwis}}, \bibinfo {author} {\bibfnamefont {M.~T.}\ \bibnamefont {Eiles}},
  \ and\ \bibinfo {author} {\bibfnamefont {A.}~\bibnamefont {Eisfeld}},\ }\href
  {\doibase 10.1103/physrevlett.124.193401} {\bibfield  {journal} {\bibinfo
  {journal} {Physical Review Letters}\ }\textbf {\bibinfo {volume} {124}},\
  \bibinfo {pages} {193401} (\bibinfo {year} {2020}{\natexlab{b}})}\BibitemShut
  {NoStop}%
\end{thebibliography}%
\end{document}